\documentclass[twocolumn, twocolappendix]{aastex631}

\usepackage{booktabs}
\usepackage{amsmath, multirow}
\usepackage{rotating}
\usepackage{longtable}

\usepackage[normalem]{ulem}     
\usepackage{soul,xcolor}        
\setstcolor{red}                
\usepackage{hyperref}           
\usepackage{threeparttable}     

\def\HI{{\rm H\,{\textsc{\romannumeral 1}}}}

\def\HIMF{{\rm H\,{\textsc{\romannumeral 1}}MF}}
\def\HIWF{{\rm H\,{\textsc{\romannumeral 1}}WF}}

\def\farcs{{$.\!\!^{\prime\prime}$}}

\graphicspath{{./}{figures/}}

\begin{document}

\title{FAST Ultra-Deep Survey (FUDS): the star formation histories of FUDS0 galaxies}

\correspondingauthor{Lister Staveley-Smith}
\email{lister.staveley-smith@uwa.edu.au}
\correspondingauthor{Bo Peng}
\email{pb@nao.cas.cn}

\author[0000-0001-6642-8307]{Hongwei Xi}
\affiliation{National Astronomical Observatories, Chinese Academy of Sciences\\
20A Datun Road, Chaoyang District, Beijing 100101, China}

\author[0000-0002-8057-0294]{Lister Staveley-Smith}
\affiliation{International Centre for Radio Astronomy Research (ICRAR), University of Western Australia,\\ 
35 Stirling Hwy, Crawley, WA 6009, Australia}

\author[0000-0001-6956-6553]{Bo Peng}
\affiliation{National Astronomical Observatories, Chinese Academy of Sciences\\
20A Datun Road, Chaoyang District, Beijing 100101, China}
\affiliation{Department of Astronomy and Institute of Interdisciplinary Studies, Hunan Normal University\\
Changsha, Hunan 410081, China}

\author[0000-0002-0196-5248]{Bi-Qing For}
\affiliation{International Centre for Radio Astronomy Research (ICRAR), University of Western Australia,\\ 
35 Stirling Hwy, Crawley, WA 6009, Australia}

\author[0000-0002-1311-8839]{Bin Liu}
\affiliation{National Astronomical Observatories, Chinese Academy of Sciences\\
20A Datun Road, Chaoyang District, Beijing 100101, China}

\author[0000-0002-7550-0187]{Dejian Ding}
\affiliation{National Astronomical Observatories, Chinese Academy of Sciences\\
20A Datun Road, Chaoyang District, Beijing 100101, China}
\affiliation{School of Astronomy and Space Science, University of Chinese Academy of Sciences\\
No.1 Yanqihu East Road, Huairou District, Beijing, 101408, China}



\begin{abstract}

    We present the ultraviolet, optical and infrared counterparts of 128 galaxies detected in neutral hydrogen (\HI) in the FAST Ultra-Deep Survey (FUDS) field 0 (FUDS0). \HI\ mass upper limits are also calculated for 134 non-detections in the field. Stellar masses ($M_*$), star formation rates (SFRs) and star formation histories are computed by fitting spectral energy distributions (SEDs) using \textsc{ProSpect}. The results show that \HI-selected galaxies prefer recent long-lasting, but mild star formation activity, while \HI\ non-detections have earlier and more intense star formation activity. Based on their distribution on the SFR versus $M_*$ diagram, the typical evolution of \HI-selected galaxies follows three distinct stages: ({\romannumeral 1}) Early stage: the total SFR increases, though the specific SFR (sSFR) decreases from 10$^{-8}$ to 10$^{-9}$ yr$^{-1}$; ({\romannumeral 2}) Mass accumulation stage: the SFR is steady, and stellar mass increase linearly with time; ({\romannumeral 3}) Quenching stage: star formation activity quenches on a rapid timescale and at constant stellar mass. 37 non-detections are located on star-forming main sequence, but are not detected in \HI\ due to low sensitivity close to field edges or close to strong radio frequency interference. Comparisons with the existing optical, optically-selected \HI, and \HI\ catalogs show a good agreement with respect to measured $M_*$ and SFR, with minor discrepancies due to selection effects. The ongoing full FUDS survey will help us better explore the evolutionary stages of \HI\ galaxies through a larger sample.


\end{abstract}

\keywords{Galaxies (573) --- Late-type galaxies (907) --- High-redshift galaxies (734) --- Galaxy evolution (594) --- Sky surveys (1464)}


\section{Introduction}\label{Sct_01}

    Hydrogen is the most abundant and basic element in the Universe, and was formed at the early stage after the Big Bang. As the Universe evolved, most of atomic hydrogen (\HI) became locked up in and around galaxies where it is dense enough to self-shield against ionizing radiation from stars and black holes. A larger fraction of the cosmic hydrogen density also remains locked up in the ionized form (H$^+$) in the intergalactic medium or has been converted to stars. \HI\ in the interstellar medium continues to regulate star formation activity by feeding star formation regions in galaxies \citep{2008A&ARv..15..189S, 2012ARA&A..50..531K}, and is therefore a key probe for studying the evolution of galaxies. Shallow \HI\ surveys such as the \HI\ Parkes All Sky Survey (HIPASS, \citealp{2001MNRAS.322..486B, 2004MNRAS.350.1195M, 2006MNRAS.371.1855W}), which detected 5,317 galaxies, and the Arecibo Legacy Fast ALFA Survey (ALFALFA, \citealp{2005AJ....130.2598G}), which detected 31,501 galaxies, have been the cornerstone for the analysis of the properties of galaxies in the local Universe. Examples of studies conducted with these datasets include: the Baryonic Tully-Fisher relation (BTFR), the \HI\ velocity width function (\HIWF), the \HI\ mass function (\HIMF), and the cosmic \HI\ density ($\Omega_\HI$) \citep{2003AJ....125.2842Z, 2005MNRAS.359L..30Z, 2010ApJ...723.1359M, 2014MNRAS.444.3559M, 2016MNRAS.457.4393J, 2018MNRAS.477....2J, 2022MNRAS.509.3268O, 2023ApJ...950...87B}.

    Beyond the local Universe, there have been several deep \HI\ surveys (some are still ongoing), including the Arecibo Ultra-Deep Survey (AUDS, \citealp{2011ApJ...727...40F, 2015MNRAS.452.3726H, 2021MNRAS.501.4550X}), the HIGHz Arecibo Survey (HIGHz, \citealp{2008ApJ...685L..13C, 2015MNRAS.446.3526C}), the Blind Ultra Deep \HI\ Environmental Survey (BUDHIES; \citealp{2007ApJ...668L...9V, 2023MNRAS.519.4279G}), the Cosmos HI Large Extragalactic Survey (CHILES; \citealp{2016ApJ...824L...1F, 2019MNRAS.484.2234H}), the Deep Investigation of Neutral Origins (DINGO; \citealp{2009PRA...........M, 2023MNRAS.518.4646R}), the MeerKAT International GHz Tiered Extragalactic Exploration (MIGHTEE-HI, \citealp{2016mks..confE...6J, 2021A&A...646A..35M}), the Looking At the Distant Universe with the MeerKAT Array (LADUMA; \citealp{2016mks..confE...4B, 2018AAS...23123107B}), and the Giant Metrewave Radio Telescope Cold-HI AT $z \approx 1$ survey (CATz1; \citealp{2022ApJ...937..103C}). However, there are only a few hundred \HI\ galaxy detections at $z>0.1$ without any firm detections reported beyond $z=0.3$ previous to FUDS. The small number and the limited redshift range of individual detections \citep{2021MNRAS.501.4550X} and stacking studies \citep{2022ApJ...941L...6C} has therefore made it difficult to factor \HI\ content into studies of the evolution of galaxies with cosmic ages $<12$ Gyr ($z>0.1$).
    
    The FAST (Five-hundred-meter  Aperture  Spherical Telescope; \citealp{2011IJMPD..20..989N}) Ultra-Deep Survey (FUDS, \citealp{2022PASA...39...19X}) is a deep blind survey aiming at detecting faint or distant \HI\ galaxies at $z<0.42$ through the 21 cm emission line to study the evolution of cool gas in galaxies. The full survey covers six independent fields evenly distributed in right ascension ($R.A.$) to minimize the cosmic variance, with an area of 0.72 deg$^2$ for each field and a total area of 4.3 deg$^2$. The  declination ($Dec$) of all fields is approximately 25$^\circ$, where the full gain of FAST is available. All  six fields are covered by the footprint of the Sloan Digital Sky Survey (SDSS), providing for detailed optical information. The field positions were chosen to avoid strong continuum sources for better spectral baselines. Thanks to the large receiving area and cryogenic 19-beam receiver \citep{8105012}, the target high sensitivity of 50\,$\mu$Jy beam$^{-1}$ at a frequency resolution of 22.9 kHz can be achieved over a wide field of view, with only $\sim 100$ hours of FAST observing time per field.
    
\begin{figure}
        \begin{center}
            \includegraphics[width=\columnwidth]{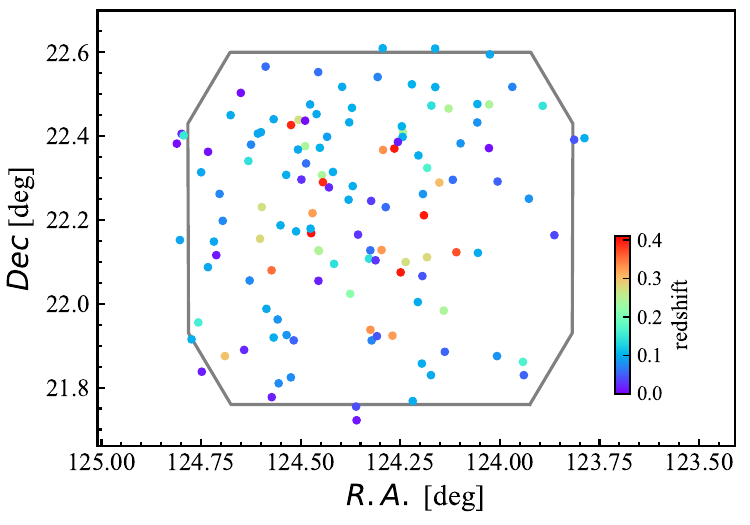}
            \caption{The spatial distribution of FUDS0 galaxies, color coded by redshift. The gray line indicates the boundary of the FUDS0 field, estimated from ideal scan lines and the geometry of the 19-beam receiver.}\label{Fig_01}
        \end{center}
    \end{figure}

    The pilot survey field, FUDS0 \citep{2024ApJS..274...18X}, is located at $(R.A., Dec) = (124.\!\!^\circ3, 22.\!\!^\circ18)$ to partially overlap with the GAL2577 field in AUDS survey for verifying the data quality. Observations were taken between 2019 August 25 and 2020 May 22. The rms noise in the final data cube is $\sim$50\,$\mu$Jy beam$^{-1}$ at a frequency resolution of 22.9 kHz around the central area in the field, with an increase toward the edges due to less sampling. In the FUDS0 survey, 128 \HI\ galaxies were detected with redshifts up to $z \sim 0.4$. Figure \ref{Fig_01} shows the sky and redshift distributions of FUDS0 galaxies.

    The FUDS0 catalog provides a unique view of cool gas in individual \HI-selected galaxies over a wider redshift span than previously possible. In combination with existing data at ultraviolet (UV), optical and infrared (IR) bands, which provide information on the stellar and dust components, a more complete picture of galaxy properties and their likely evolutionary paths can be built up for galaxies in the star-forming (SF) main sequence \citep{2004ApJ...600..681B, 2007ApJS..173..315S}. Of particular interest for studies of the evolution of galaxies are measurements of stellar mass ($M_*$), star formation rate (SFR), star formation efficiency (SFE), and star formation history (SFH).

    \citet{2020AJ....160..271D} identified the optical counterparts of ALFALFA galaxies using the SDSS catalog, and derived $M_*$ and SFR using UV, optical, and IR data. They found that ALFALFA galaxies typically have lower stellar mass and bluer colors. This is consistent with earlier work \citep{2010AJ....139..315W}, which demonstrated that the \HI-selected galaxies have lower surface brightness, fainter absolute magnitudes, bluer colors, and lower $M_*$ than those in typical SDSS (optically-selected) samples. In BTFR studies, galaxies detected by single dish telescopes \citep{2000ApJ...533L..99M, 2015MNRAS.446.3526C, 2016A&A...593A..39P} or interferometers with low resolution \citep{2019MNRAS.489.5723F, 2023MNRAS.519.4279G} crucially require optical/IR images to correct the velocity widths for inclination.

    In this paper, we derive the physical properties of FUDS0 galaxies by fitting their Spectral Energy Distributions (SEDs) in the UV, optical, and IR bands. The evolutionary status of \HI\ galaxies are investigated using their star formation histories (SFHs). The structure of the paper is described as follows. In Section \ref{Sct_02}, we describe the identification of the optical counterparts of FUDS0 galaxies. The \HI\ non-detections of SDSS sources with spectroscopic redshifts in the FUDS0 field are also cataloged and used as a reference for comparisons in Section \ref{Sct_03}. Section \ref{Sct_04} describes the photometric measurements of FUDS0 galaxies, including non-detections in UV, optical and IR bands. The processes for deriving the physical properties are detailed in Section \ref{Sct_05}. The derived SFH is used to explore the evolutionary stages of \HI\ galaxies in Section \ref{Sct_06}. Comparisons between FUDS0 and other studies are made in Section \ref{Sct_07}. Finally, we summarise our findings in Section \ref{Sct_08}. Throughout this paper, we use the flat Universe model with cosmological parameters of $H_0 = 70 \, h_{70} \, \rm km \, s^{-1} \, Mpc^{-1}$, $\Omega_{\rm M} = 0.3$ and $\Omega_\Lambda = 0.7$.

\section{Optical counterparts}\label{Sct_02}

    Due to the large beam size of single-dish telescopes, spectroscopic redshifts are essential for identifying optical counterparts, and optical images are essential for providing accurate positions ($RA$ and $Dec$). The FUDS0 field is in the footprint of the SDSS survey\footnote{\url{https://skyserver.sdss.org/}}  \citep{2000AJ....120.1579Y}. In SDSS DR15\footnote{\url{https://skyserver.sdss.org/dr15/en/home.aspx}} \citep{2019ApJS..240...23A}, the spectra of bright galaxies (with an apparent Petrosian magnitude limit of $r_{\rm P} \leq 17.77$ and a surface brightness limit of $\mu_{50} \leq 23.0$ mag arcsec$^{-2}$ at the half-light radius in $r$-band; \citealp{2002AJ....124.1810S}), luminous red galaxies ($r_{\rm P} \leq 19.5$ and color based cuts; \citealp{2001AJ....122.2267E}) are measured from 3800 to 9200 \AA\ with a wavelength resolution of $R \sim 1850$. Additionally, deeper spectroscopic observations are available from the AUDS optical catalog (AUDSOC; \citealp{2014UWAThesisH}) using the Anglo-Australian Telescope (AAT), which overlaps with the FUDS0 field. In the AUDSOC observations, the targets are 755 fainter galaxies from SDSS DR7 with $r < 20.3$ and photometric redshift $z_{\rm photo}<0.24$ in both AUDS fields (FIELD17H and GAL2577). Spectroscopic redshifts for 229 galaxies were measured with sufficient quality. Cross matching was made to identify the counterparts in DR15 by using the positions and flux densities in 5 bands. Additionally, some of the SDSS galaxies in the FUDS0 field have redshifts from Dark Energy Spectroscopic Instrument (DESI) spectra or new spectroscopic observations with Keck:I, Hale or BTA \citep{2024ApJ...966L..36X}. These redshifts are used as a supplement where no spectroscopic redshift is available in SDSS DR15. In total, there are 196 galaxies with spectroscopic redshifts in the FUDS0 field, including 143 from SDSS DR15, 46 from AUDSOC, 3 from DESI, 2 from Keck:I, 1 from Hale, and 1 from BTA \citep[for data from the last four telescopes, see][]{2024ApJ...966L..36X}.

    We used the following criteria to search for optical counterparts with spectroscopic redshifts: 1) angular distance from the \HI\ detection smaller than the half of the beam size \citep{2022PASA...39...19X}; and 2) a velocity difference less than 250 km s$^{-1}$. Note that the nearest counterpart was adopted if multiple optical sources satisfied the criteria. We identified the optical counterparts for 58 \HI\ galaxies. An exception was made for: 1)  sources located close to edge of the field, and 2) sources where RFI might have affected position accuracy. For these sources, we identified the optical counterparts for a further 4 \HI\ galaxies within a larger radius ($5'$). One was at a low frequency (1.039 GHz) and three were close to the field edge. This resulted in a total of 62 FUDS0 galaxies with counterparts in SDSS DR15 having spectroscopic redshifts.

    For the rest of the \HI\ detections, we attempted to identify their optical counterparts in DESI Legacy Imaging Survey\footnote{\url{https://www.legacysurvey.org/}} \citep{2019AJ....157..168D} DR9, which provides accurate photometric redshifts predicted using a machine learning algorithm \citep{2001MachL..45....5B}, based on the combination of optical data from the DESI Legacy Imaging Survey and IR data from Wide-field Infrared Survey Explorer (WISE).  Skyviewer\footnote{\url{https://www.legacysurvey.org/viewer}} allows us to upload a catalog with radii, which helps visualize the angular sizes of galaxies in the image. Here, we use the following criteria to identify those optical counterparts which only have photometric redshifts: 1) morphological label $TYPE$ classified as ``REX'', ``DEV'', ``EXP'' or ``SER'' to remove stars; 2) within the frequency-dependent beam \citep{2022PASA...39...19X}; 3) a redshift difference within 1-$\sigma$ uncertainty of the photometric redshift \citep{2023JCAP...11..097Z}; 4) with optical angular size close to $D_{\rm opt} = D_{\rm \HI}/1.5$, where $D_{\rm \HI}$ is estimated from the \HI\ size-mass relation \citep{2016MNRAS.460.2143W}; and 5) colors which matched those of spectroscopically-confirmed matches, as visually determined from DESI Legacy Imaging Survey pseudo-color images. This resulted in a further 56 FUDS0 galaxies with only photometric redshifts from the DESI Legacy Imaging Survey. The DESI positions were then used to find the respective SDSS counterparts, except for the counterpart to FUDS0 galaxy (ID: 069) which is not identified in SDSS DR15. In total, there are 118 FUDS0 galaxies associated with either SDSS or DESI sources. Table \ref{Tab_01} gives the $R.A.$, $Decl.$ and spectroscopic redshifts ($z_{\rm spect}$) of the optical counterparts.

    \begin{figure}
        \begin{center}
            \includegraphics[width=\columnwidth]{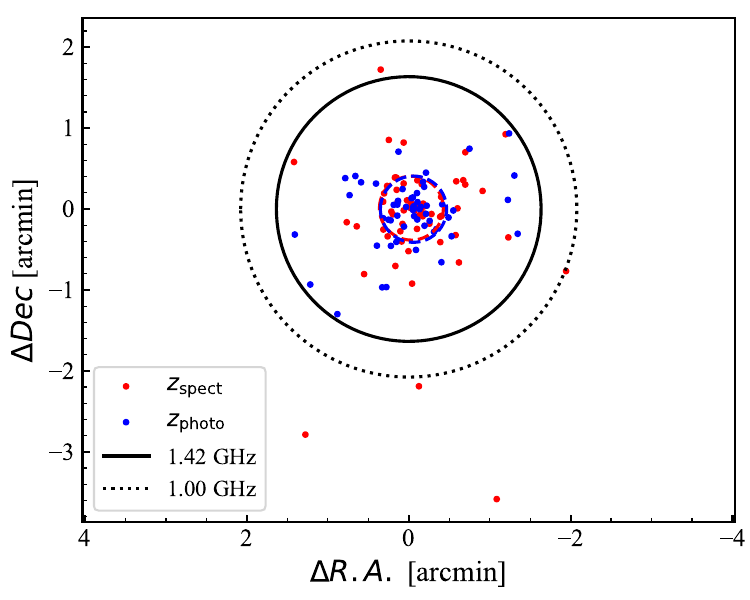}
            \caption{The 2D offset distribution of optical counterparts from the centroids of \HI\ galaxies ($\Delta R.A. = (R.A._{\rm opt} - R.A._{\HI}) \times \cos(Dec_{\HI})$ and $\Delta Dec = Dec_{\rm opt} - Dec_{\HI}$). The red dots represent the galaxies with spectroscopic redshift, while blue dots the galaxies with only photometric redshifts. The median values and 1-$\sigma$ uncertainties (dashed circles in the same colors) are calculated after removing the four outliers. The beam size at 1.0 and 1.42 GHz is shown by the black solid and dotted circles, respectively.}\label{Fig_02}
        \end{center}
    \end{figure}

\begin{figure}
        \begin{center}
            \includegraphics[width=\columnwidth]{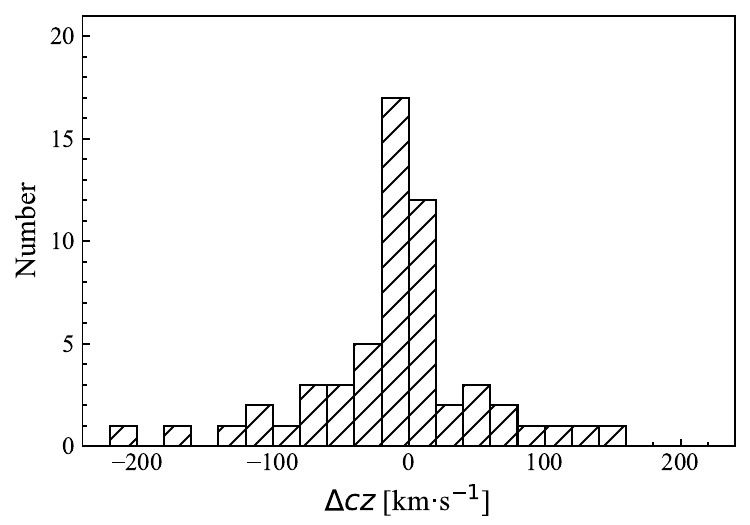}
            \caption{The distribution of $\Delta cz = c \times (z_{\rm opt, spect} - z_{\HI})$ for galaxies with spectroscopic redshift.}\label{Fig_03}
        \end{center}
    \end{figure}

\begin{figure}
        \begin{center}
            \includegraphics[width=\columnwidth]{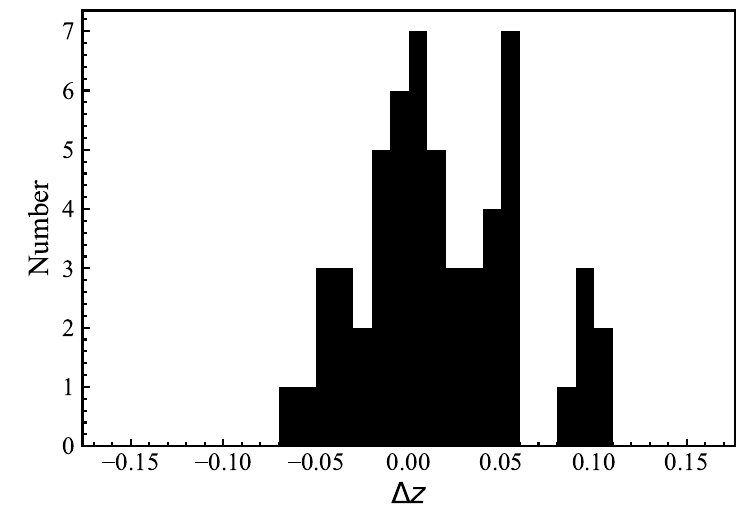}
            \caption{The distribution of redshift separation for galaxies only with photometric redshift ($\Delta z = z_{\rm opt, photo} - z_{\HI}$). Considering the accuracy, we use photometric redshift from DESI Imaging Legacy Survey.}\label{Fig_04}
        \end{center}
    \end{figure}

    In Figure \ref{Fig_02}, we show the spatial offsets of the optical counterparts from the \HI\ galaxies. The galaxies with only photometric redshifts have a similar distribution to those of galaxies with spectroscopic redshifts. The median positions and 1-$\sigma$ uncertainties (mean offset from median position divided by $\sqrt{\frac{\pi}{2}}$, derived from 2D Gaussian function) are also highly consistent with each other. The consistency confirms the accuracy of our cross matching for galaxies with only photometric redshifts. There is also no systematic offset in \HI\ positions. By using all \HI\ galaxies with optical counterparts (118 galaxies), the 1-$\sigma$ uncertainty is derived as 0.402$^\prime$. Figure \ref{Fig_03} shows the distribution of recession velocity separation for the galaxies with spectroscopic redshifts, while Figure \ref{Fig_04} shows the redshift separation for the galaxies with only photometric redshifts.

    \begin{table*}
        \centering
        \caption{The optical properties of FUDS0 galaxies. Column 1 gives the FUDS0 galaxy identifier. Column 2 contains the equatorial coordinates of the optical counterparts in J2000 epoch. The spectroscopic redshifts are given in Column 3, with 1-$\sigma$ uncertainties in parenthesis. In Column 4, we show the availability of flux measurements from FUV, NUV, $u$, $g$, $r$, $i$, $z$, $W1$, $W2$, $W3$, and $W4$ bands with T for True and F for False. We give the stellar mass, SFR, and sSFR from SED fitting by \textsc{ProSpect}, adjusted by comparing with GSWLC-2 catalog, in Column 5, 6, and 7, respectively. The 1-$\sigma$ uncertainties (given in parenthesis) for $\log(M_{*, \rm ProSpect}^{\rm adjust})$, $\log({\rm SFR_{\rm ProSpect}^{adjust}})$, and $\log({\rm sSFR_{\rm ProSpect}^{adjust}})$ are estimated using the half of differences between the 15.86 and 84.13 percentiles. For the cases where the 15.86 percentiles do not exist (negative infinity), the differences between the 50.0 and 84.13 percentiles are used. The full version in machine readable format is available online.}\label{Tab_01}
        \begin{tabular}{ccccccc}
            \toprule
            ID    & $R.A.$, $Dec$ (J2000)         & $z_{\rm spect}$ & Flag$_{\rm band}$ & $\log(M_{*, \rm ProSpect}^{\rm adjust})$  & $\log({\rm SFR_{\rm ProSpect}^{adjust}})$ & $\log({\rm sSFR_{\rm ProSpect}^{adjust}})$ \\
            FUDS0 & {\tiny HH:MM:SS.S$\pm$DD:MM:SS} & -               & -  & $\log(h_{70}^{-2} {\rm M}_\odot)$ & $\log(h_{70}^{-2} {\rm M}_\odot \, {\rm yr}^{-1})$ & $\log({\rm yr}^{-1})$   \\
            (1)   & (2)                             & (3)            & (4)                               & (5)                                                & (6)         & (7)         \\
            \midrule

            G001 & 08:18:56.45+22:22:00.2 & 0.00700  (-) &  \tt{TTTTTTTTTFF} &  7.34 (0.09) &   -2.80   (0.16) &  -10.16   (0.23) \\
            G002 & 08:17:56.79+22:26:08.6 & 0.00702  (1) &  \tt{TTTTTTTTTTT} &  8.90 (0.11) &   -1.28   (0.11) &  -10.18   (0.17) \\
            G003 & 08:18:35.61+22:30:17.3 &            - &  \tt{TTTTTTTTFTF} &  6.94 (0.11) &   -2.79   (0.11) &   -9.73   (0.09) \\
            G004 &                      - &            - &  \tt{FFFFFFFFFFF} &            - &                - &                - \\
            G005 &                      - &            - &  \tt{FFFFFFFFFFF} &            - &                - &                - \\
            G006 & 08:18:50.17+22:06:55.4 & 0.01162  (1) &  \tt{TTTTTTTTTTT} &  9.05 (0.10) &   -0.91   (0.07) &   -9.95   (0.15) \\
            G007 & 08:17:25.80+21:41:07.8 & 0.01188  (1) &  \tt{FTTTTTTTTTT} & 10.32 (0.15) &   -1.25   (1.32) &  -11.55   (1.35) \\
            G008 & 08:17:01.20+22:23:10.1 &            - &  \tt{TTTTTTTTFTF} &  7.21 (0.16) &   -2.42   (0.10) &   -9.65   (0.24) \\
            G009 & 08:19:05.10+21:47:29.0 & 0.01501  (1) &  \tt{FTTTTTTTTTT} & 10.19 (0.09) &    0.34   (0.08) &   -9.82   (0.15) \\
            G010 & 08:17:49.06+22:03:19.0 &            - &  \tt{FTTTTTTTFFF} &  7.27 (0.10) &   -2.47   (0.10) &   -9.74   (0.12) \\
             ... &                    ... &          ... &          ... &          ... &              ... &              ... \\
            \bottomrule
        \end{tabular}
        \begin{tablenotes}
            \footnotesize
            \item[1] One FUDS0 galaxy (ID: G069) only has DESI counterpart. We did not perform SED fitting for the source due to the lack of SDSS flux densities.
        \end{tablenotes}
    \end{table*}

\section{\HI\ Non-detections}\label{Sct_03}

    As mentioned in Section \ref{Sct_02}, we have 196 optical galaxies with spectroscopic redshifts in the FUDS0 field. However, only 62 of these are detected in \HI. For the 134 \HI\ non-detections, we estimate their $M_\HI$ upper limits. This requires an estimate of linewidth. Due to selection effects, many of these galaxies will be massive and have wide linewidths. We therefore factor in the linewidth information provided by the spectroscopy.

    The SDSS catalog provides information on optical emission lines for MPA/JHU galaxies \citep{2004ApJ...613..898T, 2004MNRAS.351.1151B}. Linewidths for the forbidden and Balmer transitions are measured simultaneously, then corrected for the instrumental resolution. Since the spectroscopic fibres do not contain the integrated light from the galaxy, we firstly investigate the usefulness of this approach by exploring the \HI-optical linewidth relation for the FUDS0 \HI\ detections.
     
\begin{figure}
        \begin{center}
            \includegraphics[width=\columnwidth]{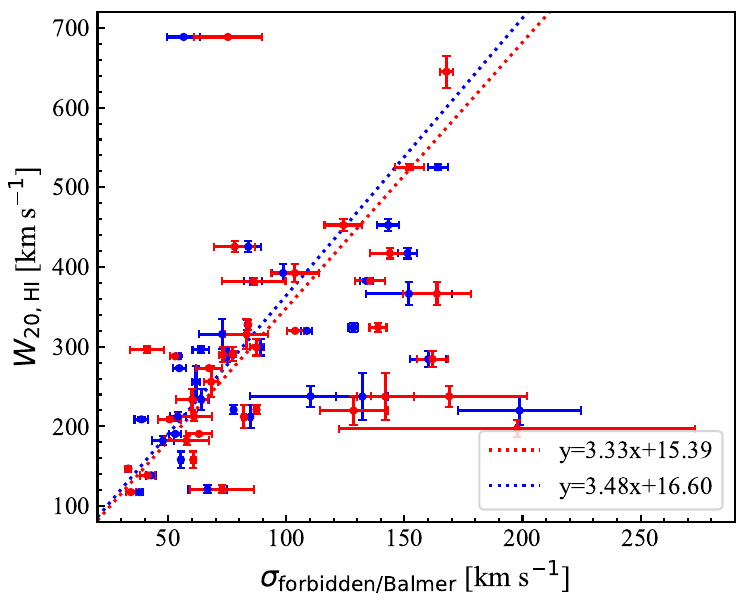}
            \caption{The \HI\ linewidth ($W_{20, \HI}$) as a function of forbidden (blue) and Balmer (red) linewidth from SDSS in the rest frame. The best fit lines are given in the same color.}\label{Fig_05}
        \end{center}
    \end{figure}

    There are 38 FUDS0 \HI\ galaxies with forbidden linewidths from SDSS, of which 36 have Balmer linewidths In Figure \ref{Fig_05}, the \HI\ linewidth ($W_{20, \HI}$) is plotted as a function of the Balmer ($\sigma_{\rm Balmer}$) and forbidden linewidth ($\sigma_{\rm  forbidden}$) in the rest frame. Despite the large scatter, the $W_{20, \HI}$ is approximately proportional to $\sigma_{\rm Balmer}$ and $\sigma_{\rm  forbidden}$. We therefore fit straight lines, using the uncertainties as weights. The best fit lines are similar:
    \begin{equation}
        W_{20, \HI}/{\rm km \, s^{-1}} = 3.33 \cdot \sigma_{\rm forbidden}/{\rm km \, s^{-1}} + 15.39 \label{Equ_05}
    \end{equation}
    \begin{equation}
        W_{20, \HI}/{\rm km \, s^{-1}} = 3.48 \cdot \sigma_{\rm Balmer}/{\rm km \, s^{-1}} + 16.60 \label{Equ_06} .
    \end{equation}

\begin{figure}
        \begin{center}
            \includegraphics[width=\columnwidth]{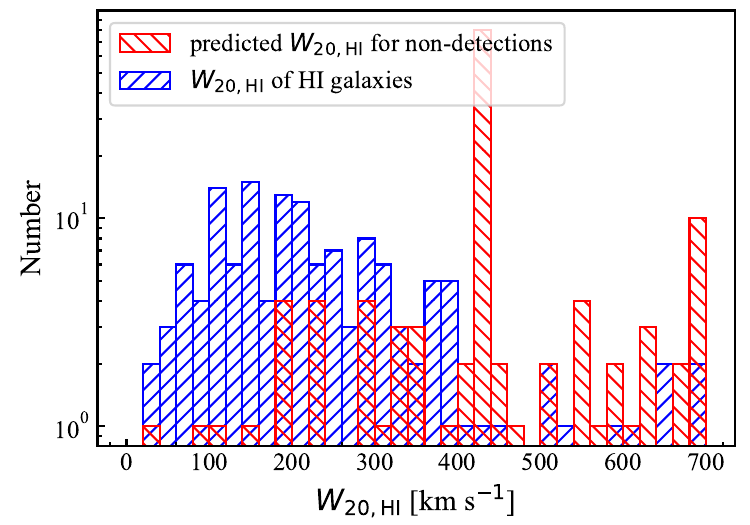}
            \caption{The distribution of \HI\ linewidth predicted by forbidden or Balmer linewidth from SDSS for the non-detections in FUDS0 field.}\label{Fig_06}
        \end{center}
    \end{figure}

    In the \HI\ non-detection catalog, there are 51 galaxies with forbidden linewidths from SDSS, along with 6 galaxies with Balmer linewidths. Equations \ref{Equ_05} and \ref{Equ_06} were used to estimate the \HI\ linewidth, prioritizing the forbidden linewidths. To avoid unrealistic values, we constrained the minimum and maximum of the predicted \HI\ linewidth to lie within the range of the FUDS0 catalog (24 -- 689 km s$^{-1}$). For the 77 galaxies without forbidden and Balmer linewidths, we assumed the median value of the predicted \HI\ linewidths (439 km~s$^{-1}$). The deviation of $W_{20, \HI}$ measurements from the best fit lines gives a 1-$\sigma$ uncertainty of 83 km s$^{-1}$ for the predicted $W_{20, \HI}$. Figure \ref{Fig_06} shows the distribution of the predicted \HI\ linewidths. Galaxies not detected in \HI\ have wider linewidth predictions, with 16 having $W_{20, \HI} > 600$ km\,s$^{-1}$ whilst there are only 5 \HI\ detections with measurements in this linewidth range. The difference is due to selection effects as SDSS spectroscopic measurements were performed on bright galaxies, which are preferentially massive galaxies with large velocity dispersions. Note that the large number in the bin at 430 km\,s$^{-1}$ corresponds to the median linewidth for the non-detections without forbidden and Balmer linewidths.

\begin{figure}
        \begin{center}
            \includegraphics[width=\columnwidth]{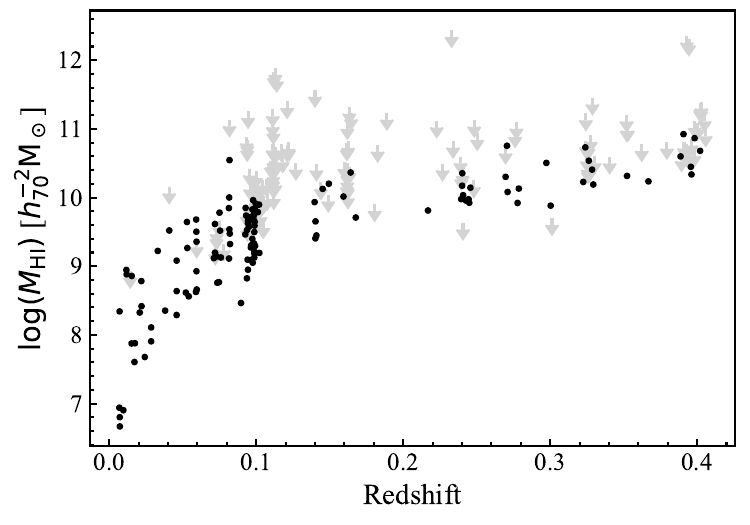}
            \caption{The upper limits of \HI\ masses as a function of redshift for non-detections (gray arrows) along with \HI\ galaxies (black dots).}\label{Fig_07}
        \end{center}
    \end{figure}

    In our previous work \citep{2024ApJS..274...18X}, we generated a noise cube with the same dimension as \HI\ data cube for the FUDS0 field. The noise in each voxel is calculated by root of mean square ($RMS$) of nearby channels in the spectrum. The median values of voxels around \HI\ non-detections (within the cube beam size in space and \HI\ linewidth predicted above in frequency) are used as the local noise for the galaxies. We also showed that the completeness of FUDS0 catalog can be expressed as a function of the noise and linewidth normalized flux, $T = \frac{S_{\rm int}}{\rm Jy \cdot Hz} \cdot \frac{\rm Jy}{\sigma} \cdot (\frac{\rm 10^6~Hz}{W_{20, \HI}})^\alpha$ (where $\alpha = 0.77$ for $W_{20, \HI} < 10^6$ Hz, and $1.56$ for $W_{20, \HI} \geq 10^6$ Hz). In the FUDS0 catalog, the completeness drops below 95\% at $T_{\rm 95} = 10^{6.344}$. We therefore calculate the flux upper limit from $\frac{S_{\rm int}}{\rm Jy Hz} < T_{\rm 95} \cdot \frac{\sigma}{\rm Jy} \cdot (\frac{W_{20, \HI}}{\rm 10^6~Hz})^\alpha$. The corresponding upper limit of \HI\ mass is therefore \citep{2017PASA...34...52M}:
    \begin{equation}
        M_{\rm \HI, uplim} = 49.7 \cdot D_{\rm Lum}^2 \cdot S_{\rm int, uplim}
    \end{equation}
    where $M_{\rm \HI, uplim}$ is the upper limit of \HI\ mass in $h_{70}^{-2} {\rm M_\odot}$, $D_{\rm Lum}$ is the luminosity distance of the galaxy in $h_{70}^{-1} {\rm Mpc}$,
    and $S_{\rm int, uplim}$ is the upper limit of integrated flux in Jy Hz. With a probability of 95\%, the non-detections have \HI\ mass lower than the predicted limits, $M_{\rm \HI, uplim}$. The median relative error of $M_{\rm \HI, uplim}$ is estimated to be 30\% by error propagation. Figure \ref{Fig_07} plots the distribution of the upper limits of \HI\ mass alongside the detections. Most of the nearby galaxies ($z<0.1$) are detected by FUDS0. Many galaxies with $0.1 < z < 0.22$ are missed due to RFI from global navigation satellite systems (GNSS). We also miss galaxies at $z>0.22$ due to sensitivity.

\section{Photometry}\label{Sct_04}

    \subsection{Ultraviolet bands}

        The Galaxy Evolution Explorer (GALEX, \citealp{2005ApJ...619L...1M}) mission is carried out by an orbiting space telescope working at UV wavelengths. The all-sky survey data is available in two broad bands: far-UV (FUV, effective wavelength of 0.154 $\mu$m); and near-UV (NUV, effective wavelength of 0.232 $\mu$m) with the Full Width Half Maximum (FWHM) of 4\farcs2 and 5\farcs3, respectively \citep{2007ApJS..173..682M}. Note that the difference between effective wavelength and emitted wavelength is a factor of $1+z$, which is up to 40\% for the highest redshift galaxies in FUDS0. We used photometric measurements in both bands, $fuv\_flux$ and $nuv\_flux$ from the GALEX General Release 6\footnote{\url{https://galex.stsci.edu/GR6/}} (GR6). The survey provides data in different survey depth (or exposure time), including All-sky Imaging Survey (AIS, $\sim 100$ seconds), Medium Imaging Survey (MIS, $\sim 1500$ seconds) and Deep Imaging Survey (DIS, only cover a small area). Note that the MIS is designed to maximize the overlapping region with SDSS. Hence, the FUDS0 field is also covered by MIS. 

        With the more accurate positions from the DESI Imaging Legacy Survey, we searched for the GALEX counterparts of FUDS0 galaxies and the \HI\ non-detections. The GALEX source within the half-light radius is identified as the counterpart. The nearest one is adopted if multiple GALEX sources are located within this radius. This results in 97 \HI\ galaxies and 60 non-detections with UV data from GALEX. All the GALEX data used in this paper can be found in MAST
        \citep{https://doi.org/10.17909/t9d30c}.

    \subsection{Optical bands}

        The SDSS\footnote{\url{https://skyserver.sdss.org/}} (\citealp{2000AJ....120.1579Y}) has mapped more than a third of the northern sky in five bands ($u$: 0.355\,$\mu$m, $g$: 0.477\,$\mu$m, $r$: 0.623\,$\mu$m, $i$: 0.762\,$\mu$m, $z$: 0.913\,$\mu$m) using the 2.5 m Sloan Foundation Telescope at Apache Point Observatory, with a FWHM resolution about 1$^{\prime\prime}$ \citep{2006AJ....131.2332G}. In Section \ref{Sct_02} and \ref{Sct_03}, we identified the optical counterparts of 117 FUDS0 galaxies and 134 \HI\ non-detections from SDSS. We adopt the model magnitude, $modelMag$, from SDSS DR15\footnote{\url{https://skyserver.sdss.org/dr15/en/home.aspx}} \citep{2019ApJS..240...23A}, which is a more reliable estimate for galaxies. Priority is given to SDSS because it provides measurements in more bands than the DESI Legacy Imaging Survey.

    \subsection{Infrared bands}

        The WISE \citep{2010AJ....140.1868W} is a space telescope launched in December 2009 aiming at scanning the sky with angular resolution of 6\farcs1, 6\farcs4, 6\farcs5 and 12\farcs0 in four IR bands ($W1$: 3.4\,$\mu$m, $W2$: 4.6\,$\mu$m, $W3$: 12\,$\mu$m, $W4$: 24\,$\mu$m). In October 2010, WISE ran out of coolant following its prime mission. The survey continued as the Near-Earth Object WISE (NEOWISE, \citealp{2014ApJ...792...30M}) at shorter wavelength bands ($W1$ and $W2$) aiming at asteroids and comets in the solar system. The unblurred WISE\footnote{\url{https://unwise.me/}} data (unWISE, \citealp{2019ApJS..240...30S}) combines all-sky WISE data and NEOWISE data to produce deeper images. Fainter objects are discovered, and more accurate positions are derived.

        The unWISE catalog provides the flux densities from forced photometry, which are extracted by applying the shape of SDSS sources on IR images in four WISE bands. The technique can measure the flux densities more accurately for faint and even non-detected sources. unWISE also provides the object identifiers in SDSS, so the sources from unWISE with same SDSS object identifier are considered as IR counterparts. As a result, we have flux density measurements in IR bands for 117 FUDS0 galaxies and 134 non-detections.
        
\begin{figure}
            \begin{center}
                \includegraphics[width=\columnwidth]{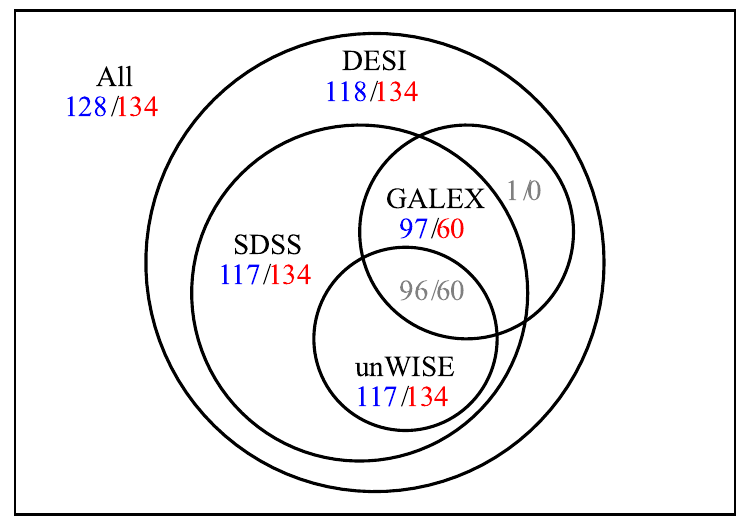}
                \caption{The galaxy numbers with counterparts from GALEX, SDSS, DESI, and WISE. The numbers for FUDS0 galaxies are given in blue, while the non-detections in red. The gray numbers indicate the galaxies in the overlap regions.}\label{Fig_08}
            \end{center}
        \end{figure}

        Figure \ref{Fig_08} shows the Venn diagram of galaxy numbers with counterparts from GALEX, SDSS, DESI, and unWISE. As noted above, FUDS0 galaxy (ID: G069) has a DESI counterpart, but no SDSS and therefore no unWISE counterpart. However, there is a GALEX counterpart.

\section{Physical properties}\label{Sct_05}

    \subsection{Milky Way extinction}

        De-reddening is needed to correct the flux densities for extinction by the Milky Way. \citet{1998ApJ...500..525S} combined data from the Cosmic Background Explorer (COBE) and the Infrared Astronomical Satellite (IRAS) to generate a full-sky dust-reddening map. The python package, \textsc{dustmaps}\footnote{\url{https://pypi.org/project/dustmaps/}} \citep{2018JOSS....3..695G} was used to derive the extinction $E(B-V)_{\rm SFD}$, and the $V$ band extinction was derived using $A_V = C_V \times E(B-V)_{\rm SFD}$, where $C_V$ is the reddening per unit $E(B-V)_{\rm SFD}$ from Table 6 in \citet{2011ApJ...737..103S}. We adopt $C_V = 2.742$ and calculate the extinction in other bands by using the dust-extinction law from \citet{1999PASP..111...63F}, using the python package, \textsc{extinction}\footnote{\url{https://extinction.readthedocs.io/en/latest/}}. The de-reddened flux densities are derived using $F_0 = F/10^{-A/2.5}$, where $F$ is reddened flux density and $A$ is the extinction at effective wavelength of the band.

    \subsection{SED fitting}

        The extent and quality of the multi-wavelength data for FUDS0 allows us to extend beyond calculations of stellar mass \citep{2011MNRAS.418.1587T, 2015ApJ...802...18M, 2014ApJ...782...90C} and SFR \citep{2011ApJ...741..124H, 2011ApJ...737...67M, 2012ARA&A..50..531K}. SED fitting (e.g. \textsc{MAGPHYS}\footnote{\url{http://www.iap.fr/magphys/}} \citealp{2008MNRAS.388.1595D}, \textsc{CIGALE} \citealp{2009A&A...507.1793N, 2019A&A...622A.103B} etc.) provides a more powerful and physical way to estimate physical parameters, and even estimate evolutionary paths that can lead to those parameters.

        In this paper, we use \textsc{ProSpect}\footnote{\url{https://github.com/asgr/ProSpect}, tutorials are provided on author's website: \url{https://rpubs.com/asgr/}} \citep{2020MNRAS.495..905R}, which is an $R$ code for SED fitting with evolving metallicities. It allows the inversion of physical parameters given broadband FUV -- FIR (0.2 -- 1000 $\mu$m) data. In addition, \textsc{ProSpect} can provide star formation histories (SFHs). The posterior distributions of the physical parameters are estimated in a Bayesian framework using the Markov Chain Monte Carlo (MCMC) method.
        
        \subsubsection{Parameter settings}

            In \textsc{ProSpect}, thirteen different functions are provided to describe the behaviour of SFH. \citet{2020MNRAS.495..905R} shows that {\tt massfunc\_snorm\_trunc}, a skewed Gaussian function with SFR suppressed at early times, better reproduces the SFHs of simulated galaxies. We adopt a modified version, in which the early SFR is forced to zero, and a base SFR was added to avoid extremely low values at late times. The maximal age, {\it magemax}, was set to the look-back time of the highest redshift galaxy at $z=14.32$ \citep{2024Natur.633..318C} relative to the look-back time of the galaxy being fit. The fit parameters in the function include {\it mSFR}, {\it mpeak}, {\it mperiod}, {\it mskew}, and {\it mbase}.

            \textsc{ProSpect} also provides three functions for evolving metallicities. We used {\tt Zfunc\_massmap\_lin} function, in which the metallicity linearly increases as the stellar mass forms. The default value ($10^{-4}$) for intial metallicity was adopted. The same maximal age was used. The fit parameter in the function is {\it Zfinal}.
            
            The spectral libraries for a simple stellar population (SSP, \citealp{2003MNRAS.344.1000B}) with a fixed initial mass function (IMF, \citealp{2003PASP..115..763C}) were used to generate the unattenuated spectra. A variant of the method in \citet{2000ApJ...539..718C} was used for dust attenuation. The library in \citet{2014ApJ...784...83D} was employed to process the re-emission in IR bands for energy balance. We did not fit the SED for an AGN component.  The fit parameters in the function are related to the dust attenuation, including {\it tau\_birth}, {\it tau\_screen}, {\it alpha\_SF\_birth}, {\it alpha\_SF\_screen}.

            The fit parameters, units, types, prior weight and ranges are detailed in Table \ref{Tab_02}. In this paper, we partially adopt the parameter ranges from \citet{2021MNRAS.505..540T}. In order to reduce the possibility of reaching the parameter space boundaries, a polynomial function was used as a prior for all parameters:
            \begin{equation}
                p(x) = - \frac{1}{100} (\frac{x-x_{\rm c}}{\sigma})^{12}
            \end{equation}
            in which $x_{\rm c} = (\max(x) + \min(x))/2$, $\sigma = (\max(x) - \min(x))/2$, and $x$ is one of the parameters in Table \ref{Tab_02}. Note that we also use the prior function for the SFR and sSFR at the age of the target galaxy in the fits, although they are not free parameters.

            \begin{table*}[!ht]
                \centering
                \caption{The units, types (in linear or logarithm scale), the prior weights ($w_{\rm prior}$), and ranges of parameters used in \textsc{ProSpect}. Note that SFR and sSFR are not free parameters. We use same prior function to constrain the ranges of SFR and sSFR at the age of the target galaxy. Hence, this information is also listed in the table.}\label{Tab_02}
                \begin{tabular}{lcccr}
                    \toprule
                    Parameter (others)      & Units                                & Type       & $w_{\rm prior}$ & Range\\
                    \midrule
                    {\it mSFR}              & $h_{70}^{-2} \rm M_\odot \, yr^{-1}$ & logarithm  & 5               & [-3, 4] \\
                    {\it mpeak}             & Gyr                                  & linear     & 5               & [-2, {\it magemax}] \\
                    {\it mperiod}           & Gyr                                  & logarithm  & 5               & [$\log(0.3)$, 2] \\
                    {\it mskew}             & -                                    & linear     & 5               & [-0.5, 1] \\
                    {\it mbase}             & $h_{70}^{-2} \rm M_\odot \, yr^{-1}$ & logarithm  & 5               & [-6, -1] \\
                    {\it Zfinal}            & -                                    & logarithm  & 5               & [-4, -1] \\
                    {\it tau\_birth}        & -                                    & logarithm  & 5               & [-2.5, 1.5] \\
                    {\it tau\_screen}       & -                                    & logarithm  & 5               & [-5, 1] \\
                    {\it alpha\_SF\_birth}  & -                                    & linear     & 5               & [0, 4] \\
                    {\it alpha\_SF\_screen} & -                                    & linear     & 5               & [0, 4] \\
                    SFR                     & $h_{70}^{-2} \rm M_\odot \, yr^{-1}$ & logarithm  & 2               & [-4, 2] \\ 
                    sSFR                    & yr$^{-1}$                            & logarithm  & 3               & [-14, -8] \\ 
                    \bottomrule
                \end{tabular}
            \end{table*}

        \subsubsection{\textsc{ProSpect} fits}

            In Section \ref{Sct_04}, we identified the counterparts of \HI\ detections with broad band flux densities from existing UV, optical or IR surveys, including GALEX, SDSS, and WISE. The available data in different bands are summarized in Table \ref{Tab_01}. SED fitting was performed for the 117 \HI\ galaxies and all 134 non-detections. One FUDS0 galaxy (ID: 069) is not included due to only one data source (FUV from GALEX) being available. Due to the lack of useful FIR data, our fits have less constraints on dust emission, and larger uncertainties. Since \textsc{ProSpect} takes into account uncertainties, we adopted a 10\%  uncertainty floor to allow for zero-point offsets \citep{2021MNRAS.505..540T}.

            \textsc{ProSpect} uses $\chi^2$ minimization (taking into acccount both likelihood and priors) to find best fit. The central values in the parameter ranges were used as initial values, then optimized by Covariance Matrix Adaptation Evolution Strategy (CMA-ES) method implemented in \textsc{cmaeshpc}\footnote{\url{https://github.com/asgr/cmaeshpc}} for $10^3$ iterations to get a more reasonable set of initial values. The Componentwise Hit-And-Run Metropolis (CHARM) algorithm in \textsc{LaplacesDemon}\footnote{\url{https://github.com/LaplacesDemonR/LaplacesDemon}} was used to perform MCMC fitting for $10^4$ interations with the intial values. Here we used \textsc{Highlander}\footnote{\url{https://github.com/asgr/Highlander}}, which combines the two steps above for convenience. The fit parameters in each iteration were recorded for further analysis.

        \subsubsection{Best fit parameters}\label{Sct_05_02_03}

\begin{figure*}
                \begin{center}
                    \includegraphics[width=2\columnwidth]{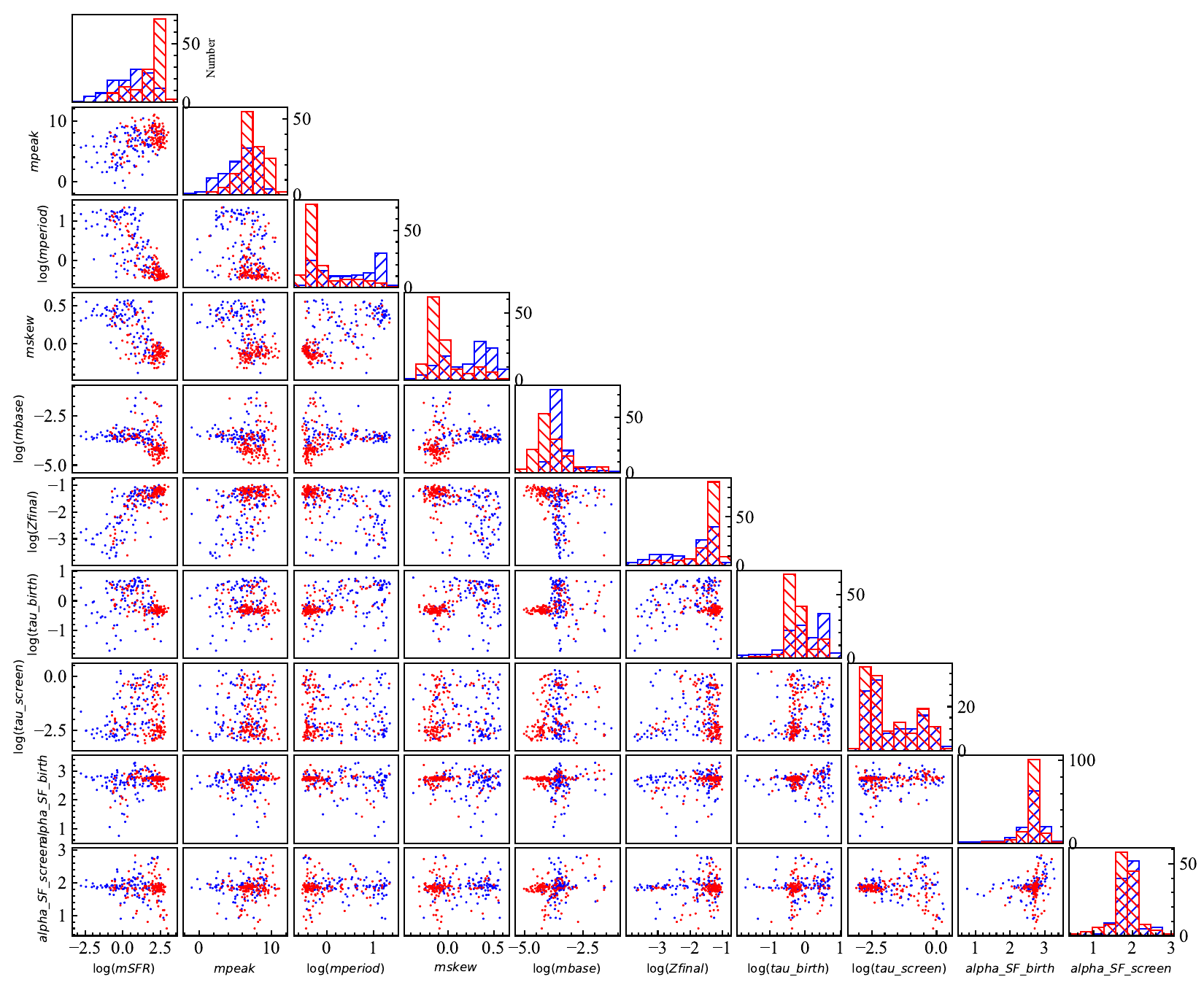}
                    \caption{The distributions of best fit parameters for both detections (blue) and non-detections (red) in FUDS0 field.}\label{Fig_09}
                \end{center}
            \end{figure*}

            Our final results consists of parameter values from $10^4$ iterations for each galaxy. We adopted the median values as the best fits. The 15.9\% and 84.1\% percentiles were used as upper and lower limits of 1-$\sigma$ uncertainties. Figure \ref{Fig_09} shows the distributions of the best fit parameters. We can see that most the parameters lie away from the boundaries, confirming the the initial parameter ranges. Some galaxies have metallicities close to the upper limit. However, most of these are \HI\ non-detections which indicate evolved and quenched galaxies.

            SFH is characterized by five parameters: peak SFR ({\it mSFR}); the look-back time of peak SFR from the galaxy redshift ({\it mpeak}); the length of time spent at high SFR ({\it mperiod}); the skewness parameter ({\it mskew}), where positive values indicate that the SFR increases quickly then decreases slowly, and negative values indicate the opposite; and the non-zero floor of SFR as the fraction of {\it mSFR} after {\it mpeak} ({\it mbase}).
            
\begin{figure}
                \begin{center}
                    \includegraphics[width=\columnwidth]{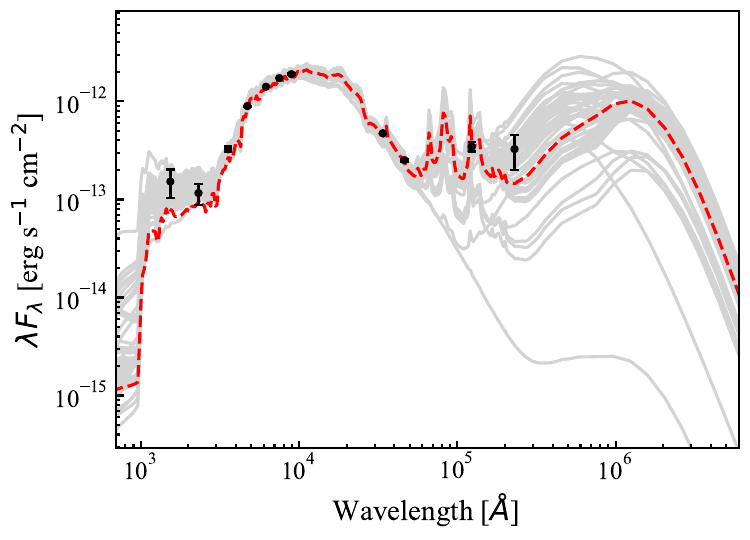}
                    \caption{An example SED for a FUDS0 galaxy (ID: G039) in the observer frame. The observation data points are represented by black dots. The red dashed line is the best fit, while the gray lines are the 50 posterior sampling randomly selected from 10$^4$.}\label{Fig_10}
                \end{center}
            \end{figure}

\begin{figure}
                \begin{center}
                    \includegraphics[width=\columnwidth]{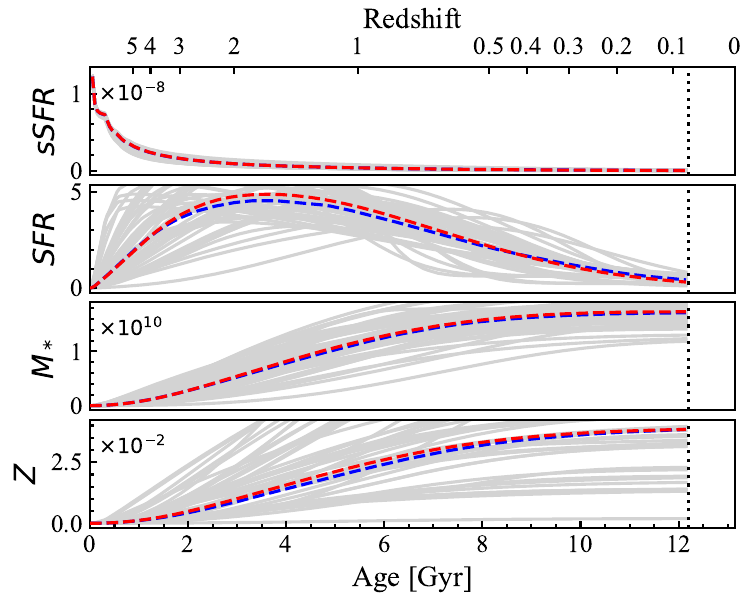}
                    \caption{The sSFR, SFR, stellar mass, and metallicity as functions of time for a FUDS0 galaxy (ID: G039). The red dashed lines are the best fit SFH. The gray lines are the same 50 posterior sampling shown in Figure \ref{Fig_10}, while the blue dashed lines are the median values. The black dotted lines indicate the redshift of the galaxy.}\label{Fig_11}
                \end{center}
            \end{figure}

            Figure \ref{Fig_09} shows the distribution in parameter space for both FUDS0 detections and non-detections. The detections mainly appear to have long lasting (high {\it mperiod}), later (smaller {\it mpeak}), milder SFR (lower {\it mSFR}), and higher SFR floor (higher {\it mbase}) than non-detections. As the non-detections would suggest, these systems have much higher metallicity, and earlier and stronger star formation bursts. Figure \ref{Fig_10} shows an example SED for FUDS0 galaxy (ID: G039). The evolution of sSFR, SFR, stellar mass, and metallicity is shown in Figure \ref{Fig_11}.

    \subsection{Comparison with GSWLC-2}\label{Sct_05_02_04}

        The GALEX-SDSS-WISE Legacy Catalog\footnote{\url{https://salims.pages.iu.edu/gswlc/}} (GSWLC) contains physical properties ($M_*$, dust attenuation and SFR) of galaxies from SDSS DR10 with spectroscopic redshifts $z \leq 0.3$, $r$-band Petrosian magnitude $r_{\rm P} \leq 18.0$, and covered by the GALEX survey (GR 6/7). The first version, GSWLC-1 \citep{2016ApJS..227....2S} contains stellar mass and SFR from fits to UV and optical SEDs using {\it Code Investigating GALaxy Emission}\footnote{\url{https://cigale.lam.fr/}}' (CIGALE, \citealp{2009A&A...507.1793N, 2019A&A...622A.103B}) along with SFRs from WISE 22\,$\mu$m photometry. The updated version, GSWLC-2 \citep{2018ApJ...859...11S} has more accurate SFRs by incorporating WISE mid-IR (MIR) flux densities and utilizing flexible attenuation curves in SED fitting. Three sub-versions are provided in GSWLC-2: A, M, and D, depending on the depth of GALEX survey. The X version combines the deepest GALEX photometry from A, M, and D, and contains $\sim$660,000 galaxies with redshift between 0.01 and 0.3.

        We adopted the latest and largest version, GSWLC-X2. However, the distances used assume a flat universe with WMAP7 cosmological parameters: $H_0 = 70.4 \, \rm km \, s^{-1} \, Mpc^{-1}$, $\Omega_{\rm M} = 0.272$. We therefore adjusted their $M_*$ and SFR measurements to match the cosmological parameters adopted in this paper. Since GSWLC-2 covers 90\% of the SDSS sky, we use it as our reference catalog for adjusting stellar mass, SFR and sSFR from \textsc{ProSpect}, so as to avoid any systematic bias when comparing different catalogs. We cross matched the SDSS counterparts of FUDS0 galaxies and \HI\ non-detections with GSWLC-2. As a result, 38 FUDS0 galaxies and 57 non-detections have GSWLC-2 counterparts.

\begin{figure}
            \begin{center}
                \includegraphics[width=\columnwidth]{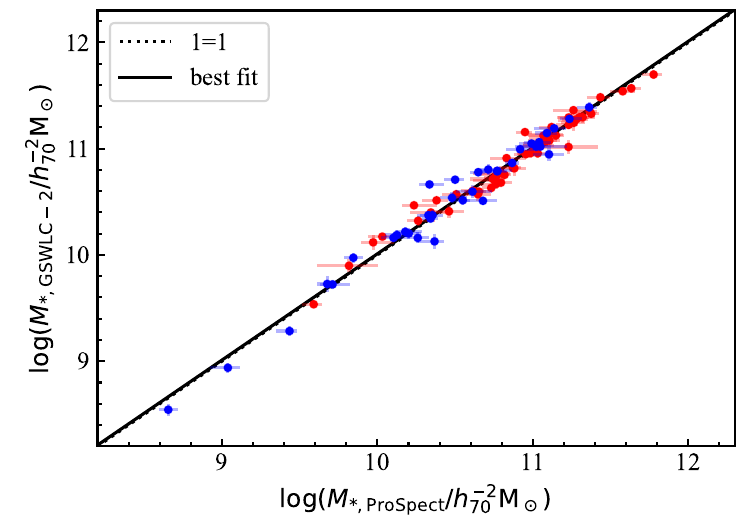}
                \caption{The comparison of stellar mass between GSWLC-2 and \textsc{ProSpect}. The blue dots are the FUDS0 \HI\ galaxies, while red dots are the non-detections in the FUDS0 field. The dotted line indicates equality, while the solid line is the best fit.}\label{Fig_12}
            \end{center}
        \end{figure}

        In Figure \ref{Fig_12}, we compare the stellar masses from \textsc{ProSpect} and GSWLC-2. The data points show a tight linear correlation along a line of unit slope. After weighting with the estimated uncertainties, the best fit offset is:
        \begin{equation}
            \begin{split}
                & \log(M_{*, \rm GSWLC-2}/ h_{70}^{-2} \rm M_\odot)\\
                &= \log(M_{*, \rm \textsc{ProSpect}}/ h_{70}^{-2} \rm M_\odot) - 0.0104 \label{Equ_09} .\\
            \end{split}
        \end{equation}
        We use the Equation \ref{Equ_09} to adjust the stellar masses from \textsc{ProSpect}.

\begin{figure}
            \begin{center}
                \includegraphics[width=\columnwidth]{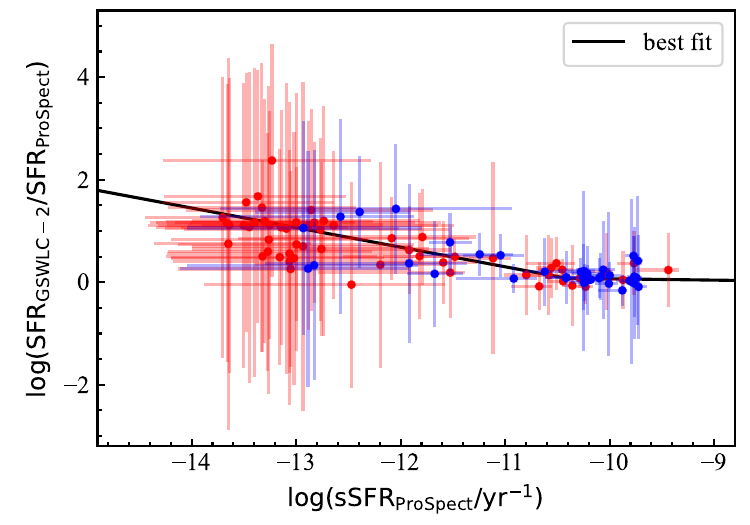}
                \caption{The ratio of SFR from GSWLC-2 and \textsc{ProSpect} as a function of sSFR from \textsc{ProSpect}. The blue symbols represent the FUDS0 \HI\ galaxies, while the red symbols the non-detections. The best fit is indicated by black solid line.}\label{Fig_13}
            \end{center}
        \end{figure}

\begin{figure}
            \begin{center}
                \includegraphics[width=\columnwidth]{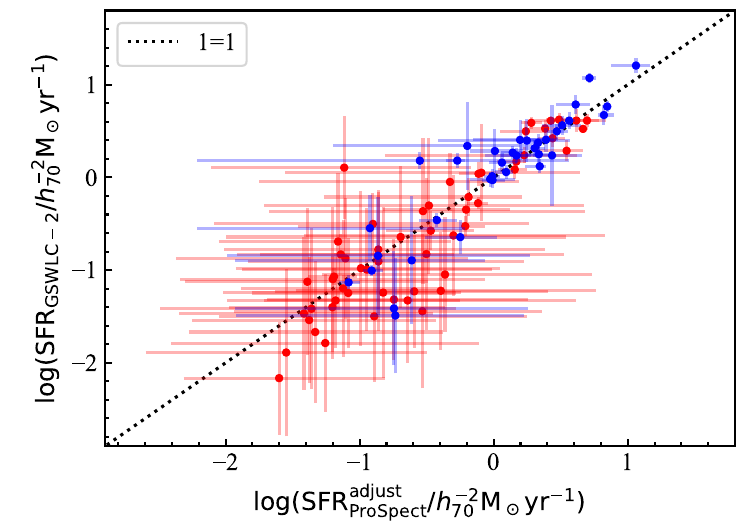}
                \caption{The comparison of SFR between GSWLC-2 and \textsc{ProSpect} after adjustment. The blue dots are the FUDS0 \HI\ galaxies, while red dots are the non-detections in the FUDS0 field. The 1 to 1 line is given by black dotted line.}\label{Fig_14}
            \end{center}
        \end{figure}

        Figure \ref{Fig_13} shows the ratio of SFR from the two methods as a function of sSFR. For the galaxies with high sSFR, the results from two methods shows consistency. However, the difference is more obvious towards low sSFR end. We use two joint lines to fit the data. The best fit lines are given below:
        \begin{equation}
            \begin{split}
                & \log({\rm SFR}_{\rm GSWLC-2}/{\rm SFR}_{\rm \textsc{ProSpect}}) \\
                & =\begin{cases}
                    -0.3822(\log({\rm sSFR}_{\rm \textsc{ProSpect}}/ \rm yr^{-1}) - 10.4170) + 0.0780  \\
                    ({\rm for}~\log({\rm sSFR}_{\rm \textsc{ProSpect}}/ \rm yr^{-1}) \leq 10.4170) \\
                    -0.0265(\log({\rm sSFR}_{\rm \textsc{ProSpect}}/ \rm yr^{-1}) - 10.4170) + 0.0780 \\
                    ({\rm for}~\log({\rm sSFR}_{\rm \textsc{ProSpect}}/ \rm yr^{-1}) > 10.4170) .\\
                \end{cases}
            \end{split} \label{Equ_10}
        \end{equation}
        We use Equation \ref{Equ_10} to adjust the SFR from \textsc{ProSpect}. Figure \ref{Fig_14} shows the correlation of SFR between two methods after the adjustment, where there is no residual offset. Note that the SFH with adjusted parameters will produce a different stellar mass. Hence, we only use the equations to adjust the SFR as a function of time. We also use both Equation \ref{Equ_09} and \ref{Equ_10} to adjust the specific SFR (sSFR) from \textsc{ProSpect}. Hereafter, we will use the adjusted stellar mass, SFR, and sSFR for FUDS0 galaxies and non-detections to keep consistency with GSWLC-2.
            
    \subsection{Final results}
        
\begin{figure}
            \begin{center}
                \includegraphics[width=\columnwidth]{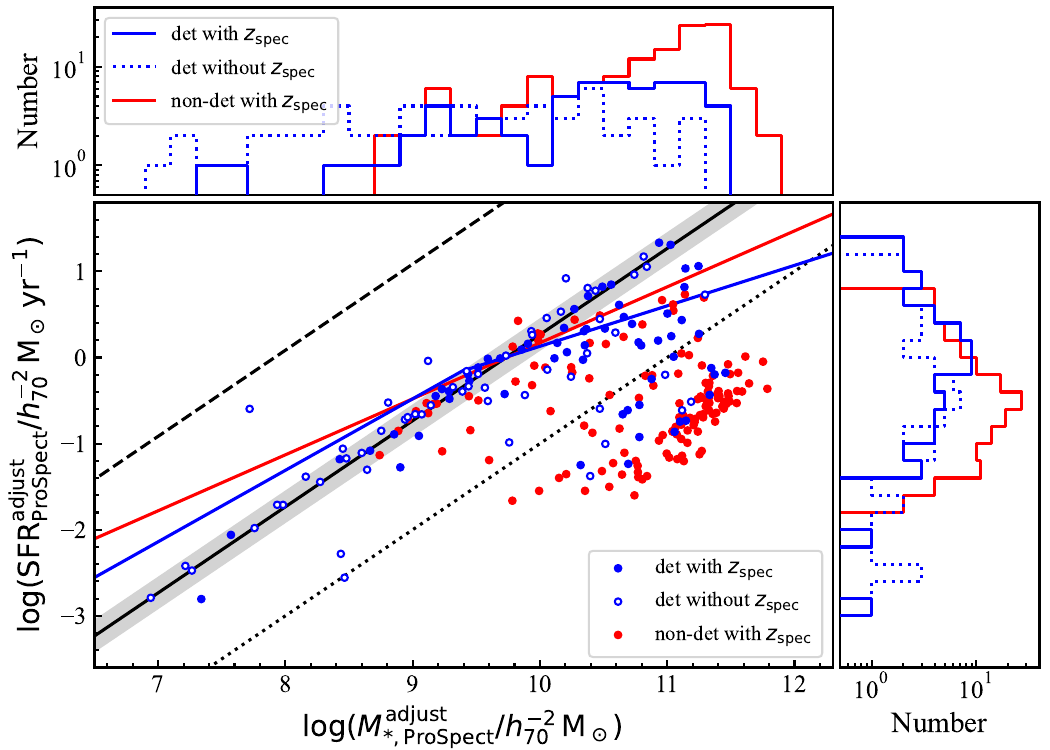}
                \caption{The distribution of detections (blue dots) and non-detections (red dots) in the SFR -- $M_*$ plane. The solid dots indicate the galaxies with spectroscopic redshifts, while the empty dots indicate those without. The black solid line represents the SF sequence with the gray area indicating the 1-$\sigma$ uncertainty derived in this paper, while the black dotted line indicates the criterion for selecting SF galaxies ($\log({\rm SFR / yr^{-1}}) > -11$). The black dashed line indicates the upper limit for peak sSFR derived from the SFHs of FUDS0 galaxies. The red and blue solid lines are the SF lines for SF or SF/AGN galaxies from \citet{2007ApJS..173..267S}.}\label{Fig_15}
            \end{center}
        \end{figure}

        The adjusted values for $\log(M_*)$, $\log({\rm SFR})$ and $\log({\rm sSFR})$ are given in Table \ref{Tab_01}. Figure \ref{Fig_15} shows the distribution of the 117 FUDS0 galaxies and 134 non-detections in the adjusted SFR versus adjusted stellar mass plane. We can see that most of FUDS0 sources are SF galaxies lying above the black dotted line defined by $\log({\rm sSFR / yr^{-1}}) > -11$ \citep{2018ApJ...859...11S}, whilst most of the non-detections are located below this in the green valley or red sequence. Such a distribution is consistent with the inference in Section \ref{Sct_05_02_03}. In Appendix \ref{Sct_A}, we examine the impact of noise, and find that 37 non-detections could still be \HI-rich galaxies missed by FUDS0 due to low sensitivity at frequencies near RFI or regions close to the field edge.
            
        Figure \ref{Fig_15} shows that the SF \HI\ galaxies distribute along a line of constant sSFR, with larger scatter at high stellar masses. To minimize the impact of outliers, the SF sequence line was derived using the median sSFR, with the `1-$\sigma$' uncertainty replaced with $1.4826 \times MAD$ (Median Absolute Deviation), giving $\log({\rm sSFR} / {\rm yr}^{-1}) = -9.73 \pm 0.18$. The equivalent for the line in SFR -- $M_*$ plane is:
        \begin{equation}
            \log({\rm SFR} / h_{70}^{-2} {\rm M}_\odot {\rm yr}^{-1}) = \log(M_* / h_{70}^{-2} {\rm M}_\odot) - 9.7338.
        \end{equation}
        The fitted line is slightly higher at the high stellar masses ($\log(M_*/h_{70}^{-2} \rm M_\odot) > 9.76$), and lower at low stellar masses compared with the fits of \citet{2007ApJS..173..267S} for SF and SF/AGN galaxies, also shown in Figure \ref{Fig_15}. We also computed the SF sequence lines for low and high redshift bins split at $z=0.095$, containing 47 and 48 SF galaxies, respectively. The results are $-9.74 \pm 0.13$ and $-9.71 \pm 0.16$, which are consistent within 1-$\sigma$ uncertainty. Although FUDS0 galaxies covers redshift up to $z \sim 0.4$, their distribution in $M_*$ -- SFR plane shows only weak evolution.

        Figure \ref{Fig_15} also shows the corresponding stellar mass and SFR histograms of \HI\ galaxies with and without spectroscopic redshifts ($z_{\rm opt, spect}$), as well as non-detections with spectroscopic redshifts. The \HI-selected galaxies (solid blue histograms) prefer lower stellar mass but higher SFR compared with non-detections (solid red histograms). For the \HI\ detections, the galaxies with spectroscopic redshifts (which are optically brighter in general) have both higher stellar mass and SFR compared with \HI\ detections without (blue dotted histograms). Our findings are consistent with \citet{2010AJ....139..315W}, who found that \HI-selected galaxies have lower stellar mass than SDSS sources. \citet{2020AJ....160..271D} found a similar offset for bright galaxies in \HI\ samples by comparing galaxies in both ALFALFA and GSWLC-2 with those only in ALFALFA. They found a slightly higher SFR but similar stellar mass for \HI-selected galaxies in the bright galaxy sample by comparing galaxies in both ALFALFA and GSWLC-2 with those only in GSWLC-2.

\section{Evolutionary stages of \HI\ galaxies}\label{Sct_06}

\begin{figure}
        \begin{center}
            \includegraphics[width=\columnwidth]{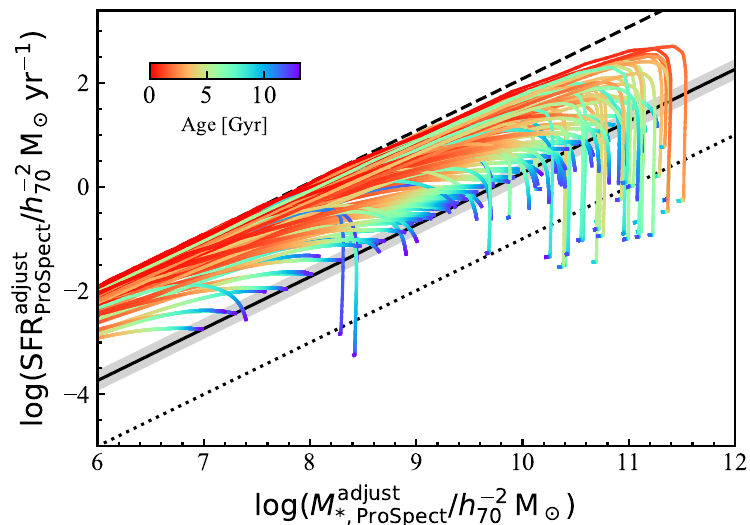}
            \includegraphics[width=\columnwidth]{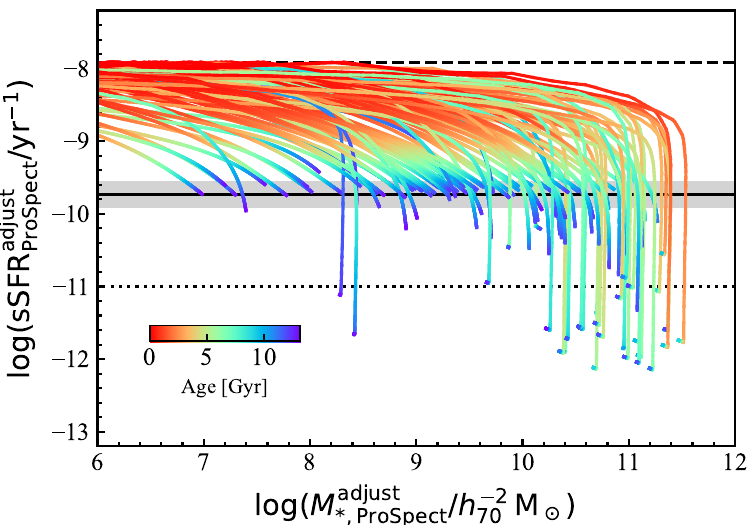}
            \caption{The best fit evolutionary tracks of \HI\ galaxies in $\log({\rm SFR})$ -- $\log(M_*)$ (upper panel) and $\log({\rm sSFR})$ -- $\log(M_*)$ (lower panel). The colors represent the ages of galaxies. The black lines have the same meanings as that in Figure \ref{Fig_15}.}\label{Fig_16}
        \end{center}
    \end{figure}

    \textsc{ProSpect} provides stellar mass, SFR and sSFR as functions of age\footnote{In this paper, age indicates the time past since $z=14.32$. It is not the age used in \textsc{ProSpect} code, which represents the look-back time at the galaxy's redshift.}. It allows us to inspect the past evolutionary tracks of galaxies in the $\log({\rm SFR})$ -- $\log(M_*)$ and $\log({\rm sSFR})$ -- $\log(M_*)$ diagrams. The stellar mass, SFR and sSFR were derived as functions of age using the best fit parameters, then adjusted using Equation \ref{Equ_09} and \ref{Equ_10}. Figure \ref{Fig_16} shows the evolutionary tracks for FUDS0 galaxies from which the following conclusions can be inferred: 
    \begin{enumerate}
        \item Early stage: at early times, stellar feedback does not limit growth, so SFR increases as mass accumulates. All FUDS0 galaxies follow a similar trend whereby sSFR decreases slowly from $10^{-8}$ to $10^{-9}$ yr$^{-1}$. The upper limit to sSFR for all FUDS0 galaxies during their evolution is 10$^{-7.9}$ yr$^{-1}$.
        
        \item Mass accumulation stage: sSFR gradually reduces, and at approximately $10^{-9}$ yr$^{-1}$, FUDS0 galaxies achieve an approximately constant SFR, and so continue to accumulate stellar mass for a long time. This is the reason that most lie along the SF main sequence. A few, mostly massive, galaxies stay in this stage for only a short time.
        
        \item Quenching stage: when the galaxy exhausts its gas, the SFR drops quickly enough that the stellar mass remains nearly unchanged. The galaxy stays with a low SFR, only supported by ambient gas accretion and mergers. Note that some galaxies which are in this stage still have $\log({\rm sSFR}/{\rm yr^{-1}}) > -11$, so are actually interlopers in the SF sequence.
    \end{enumerate}

    The above evolutionary scenario suggests that the simple picture of \citet{2007ApJS..173..315S} that steady star formation can carry a galaxy along the SF sequence is only partially correct, even for the most massive galaxies, and that sSFR does not remain constant. However, an important caveat is that we have only examined the evolution of galaxies with simple \textsc{ProSpect} models, the SFH of FUDS0 galaxies may in fact be even more complex, with galaxies rapidly changing position in the SFR -- $M_*$ plane due to mergers and interactions.

    \begin{figure}
        \begin{center}
            \includegraphics[width=\columnwidth]{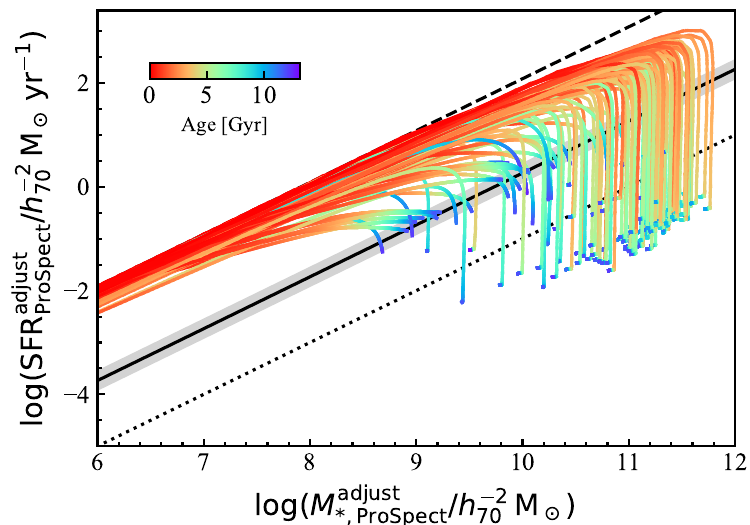}
            \includegraphics[width=\columnwidth]{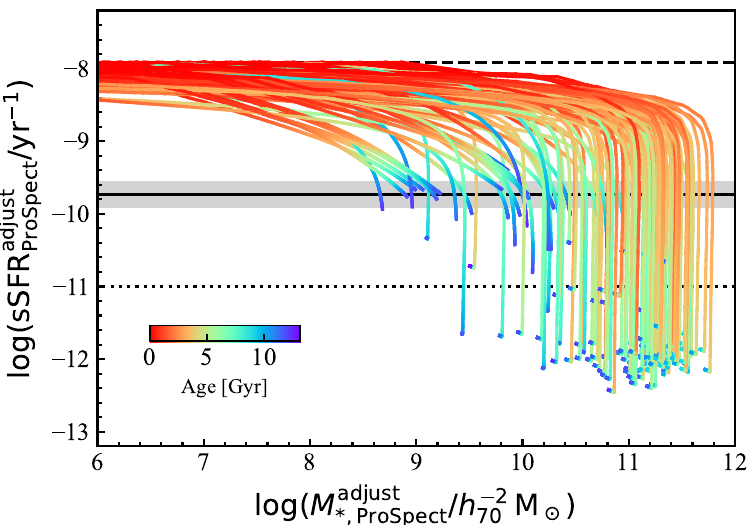}
            \caption{Same as Figure \ref{Fig_16}, but for non-detections.}\label{Fig_17}
        \end{center}
    \end{figure}
    
    Most FUDS0 galaxies will be able to maintain themselves for many Gyr in the Mass accumulation stage (also see the distribution of {\it mperiod} in Figure \ref{Fig_09}). However, some of the high stellar mass ($\log(M_*) > 10$) galaxies will exhaust their gas very quickly, so that they only stay in the Mass accumulation stage for a short time. The galaxies with high stellar mass will have undergone multiple major mergers, maintaining star formation activity and keeping them in the Early stage for a long time. However, as merger and accretion rates diminish, there is insufficient gas to sustain their enhanced SFR. Therefore, these galaxies only stay in the Mass accumulation stage for a short time before entering the Quenching stage and entering the green valley or red sequence.

    Appendix \ref{Sct_B} shows the distribution of \HI\ galaxies and non-detections in the SFR -- $M_*$ and sSFR -- $M_*$ planes at $z=3, 2, 1, 0$. They clearly show a tight SF sequence evolving as redshift for FUDS0 galaxies, i.e. a higher SF sequence line for galaxies at higher redshift. However, it is not obvious for non-detections. Considering the small fraction of high redshift galaxies ($z>0.3$) and the slow evolution of the SF sequence line at $z<0.4$, the impact of evolution is negligible for the SF sequence line calculation in this paper.

    Figure \ref{Fig_17} shows the evolutionary tracks of the 134 non-detections, also showing the three evolutionary stages. However, for the non-detections, most of the galaxies have now entered the Quenching stage. Such a result is consistent with our knowledge of the role of \HI\ gas. Similar to the high stellar mass \HI\ galaxies, these galaxies also have a short Mass accumulation stage. Some of the galaxies with intermediate stellar masses are still in the Mass accumulation stage, located near the SF sequence line. But these could be the \HI\ galaxies missed by FUDS0 due to the non-uniform noise level in the field (see Appendix Section \ref{Sct_A}).

\section{Comparison with other catalogs}\label{Sct_07}

    In this section, we compare FUDS0 galaxies with other recent catalogs, including an optical catalog (GSWLC-2), two optically-selected \HI\ catalogs (xGASS and HIGHz), and an \HI-selected catalog (ALFALFA). Note that FUDS0 is a pilot survey which consists of a limited number of galaxies compared with larger sample that will be available with full FUDS, so these results are preliminary.

    \subsection{GSWLC-2}

\begin{figure}
            \begin{center}
                \includegraphics[width=\columnwidth]{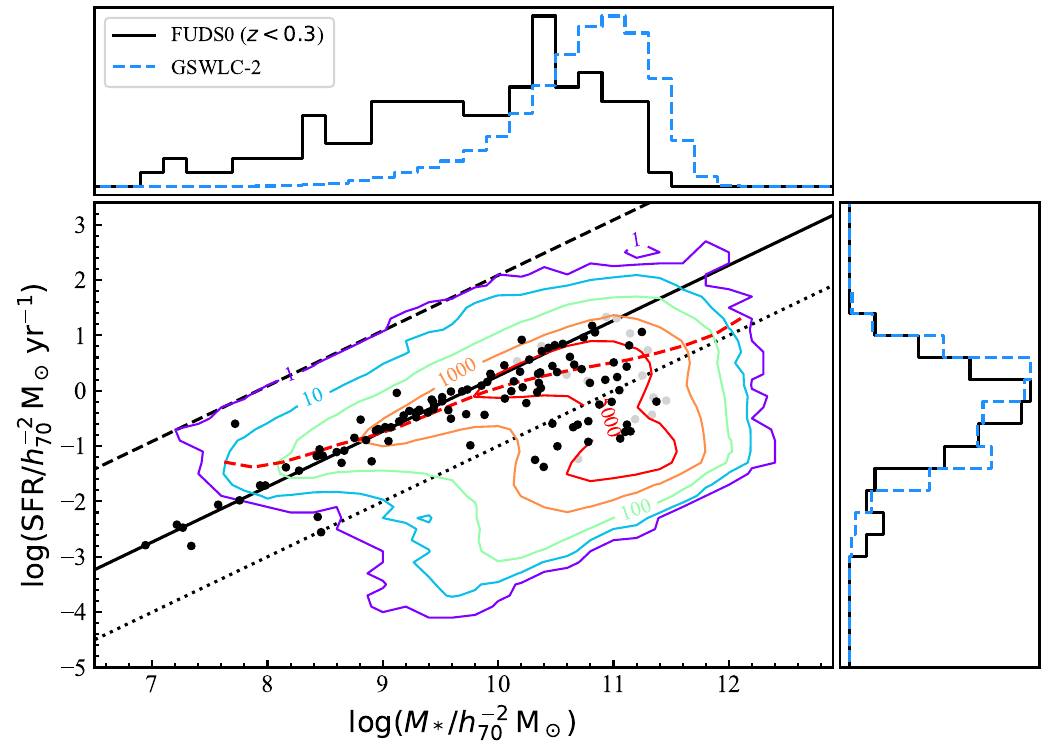}
                \caption{Comparison between FUDS0 and GSWLC-2. The distribution of GSWLC-2 galaxies is shown with coloured contours. The dashed red line shows the median SFR versus stellar mass for SF galaxies in GSWLC-2. The black dots indicate the FUDS0 \HI\ galaxies in same redshift range as GSWLC-2 ($z<0.3$), while the light gray dots the rest of the FUDS0 detections. The distributions of stellar mass (upper panel) and SFR (right panel) are normalized.}\label{Fig_18}
            \end{center}
        \end{figure}

        GSWLC-2 is an optical catalog, already detailed in Section \ref{Sct_05_02_04}. Figure \ref{Fig_18} compares FUDS0 galaxies with GSWLC-2 galaxies. Since FUDS0 has a wider redshift coverage, we only use the galaxies in the same redshift range ($z < 0.3$). There are 659,220 GSWLC-2 galaxies in the redshift range, for which 650,591 have both stellar mass and SFR measurements from CIGALE. There are 119 FUDS0 galaxies with $z < 0.3$, out of which 103 have adjusted stellar masses and SFRs from \textsc{ProSpect}. Both catalogs have similar distribution in SFR. However, FUDS0 galaxies have lower stellar masses, and lack \HI\ counterparts at the high-mass end. \citet{2010AJ....139..315W} also show that \HI-selected galaxies have lower stellar masses than SDSS sources. The discrepancy at high masses is due to the inclusion of gas-poor galaxies in the green valley or red sequence in the GSWLC-2 catalog. The difference disappears in comparison with xGASS or ALFALFA, so is due to different evolutionary stages being sampled in optically-selected samples.
        
        Most of the FUDS0 \HI\ galaxies trace the ridge of GSWLC-2 distribution. The median SFR versus stellar mass for GSWLC-2 galaxies in the SF sequence ($\log({\rm sSFR}/{\rm yr^{-1}})>-11$; \citealp{2018ApJ...859...11S}) is shown by the red dashed line. The GSWLC-2 median line matches FUDS0 galaxies in the SF sequence between $M_*=10^{8.5}$ M$_{\odot}$ and $10^{10}$ M$_{\odot}$, showing that HI and optical selection produces similar SF sequences. However, above $M_*=10^{10}$ M$_{\odot}$ the GSWLC-2 median lies below the SF sequence fit derived in this paper. As we discussed in Section \ref{Sct_06}, some Quenching stage galaxies (defined in Section \ref{Sct_06}) may be mis-classified as SF galaxies by the criterion of $\log({\rm sSFR}/{\rm yr^{-1}})>-11$. The small discrepancy at $M_*<10^{8.5}$ M$_{\odot}$ may be caused by a greater fraction of Early stage galaxies (defined in Section \ref{Sct_06}) in GSWLC-2.

\begin{figure}
            \begin{center}
                \includegraphics[width=\columnwidth]{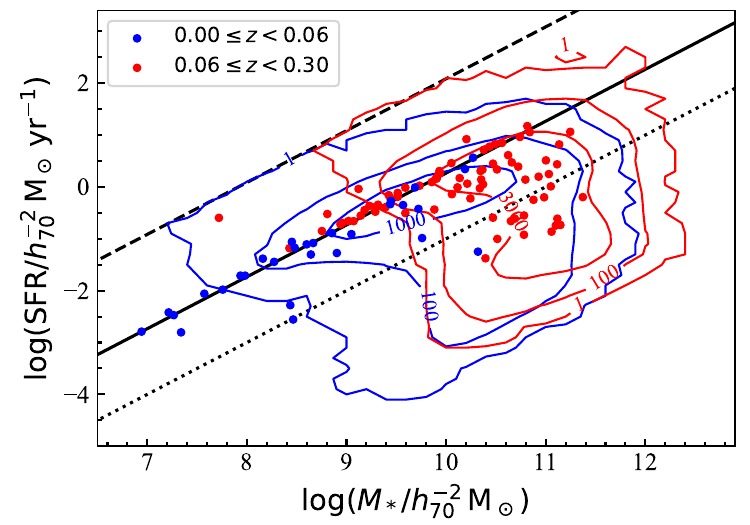}
                \caption{The distributions of GSWLC-2 (contours) and FUDS0 (dots) galaxies. The colors indicate galaxies in different redshift bins (blue: 0 -- 0.06, red: 0.06 -- 0.3). The black lines have same meanings as that in Figure \ref{Fig_15}}\label{Fig_19}
            \end{center}
        \end{figure}

        Figure \ref{Fig_19} shows the distributions of FUDS0 and GSWLC-2 galaxies in two redshift bins, $0< z \leq 0.06$ and $0.06 < z \leq 0.3$. In the low redshift bin, there are 35 FUDS0 galaxies, of which 29 have stellar masses and SFRs. FUDS0 contains more low stellar mass galaxies than GSWLC-2. However, due to the small survey field, FUDS0 lacks high stellar mass galaxies. At the low-mass end ($\log(M_*/h_{70}^{-2} \rm M_\odot)<9.5$), FUDS0 galaxies trace the ridge of the GSWLC-2 distribution. At the high-mass end ($\log(M_*/h_{70}^{-2} \rm M_\odot) \geq 9.5$), there are only 7 low-redshift galaxies, so comparison is difficult.
        
        In the high redshift bin, there are 84 FUDS0 galaxies, of which 74 galaxies have stellar masses and SFRs. There is a good consistency between the two catalogs for low-mass galaxies. FUDS0 galaxies trace the ridge of GSWLC-2 distribution, and extend to lower stellar mass. These low-mass galaxies also trace the ridge of GSWLC-2 distribution in the low redshift bin, indicating weak evolution in the SF sequence line. Meanwhile, high-mass FUDS0 galaxies show larger scatter, similar to the GSWLC-2 distribution. GSWLC-2 has more galaxies in the green valley and red sequence, most of which are of high stellar mass. We infer that the difference at high-mass end between full FUDS0 sample and full GSWLC-2 sample is caused by sample selection and more rapid evolution of high-mass optically-selected galaxies.

    \subsection{xGASS}\label{Sct_07_02}

        The extended GALEX Arecibo SDSS Survey (xGASS, \citealp{2018MNRAS.476..875C}) is an \HI\ survey (from either archive data or new observations with the Arecibo telescope) for studying gas fraction scaling relations. The target was a stellar-mass selected sample of SDSS sources with UV emission from GALEX. \HI\ observations were taken down to a gas fraction limit of $\sim 2\%$. The survey consists of two parts: 1) the GALEX Arecibo SDSS survey (GASS, \citealp{2010MNRAS.403..683C}) containing massive system with $\log(M_* / h_{70}^{-2} {\rm M_\odot}) > 10$ from SDSS DR6 \citep{2008ApJS..175..297A} and 2) the low-mass extension of GASS (GASS-low) containing SF galaxies with $9 < \log(M_* / h_{70}^{-2} {\rm M_\odot}) < 10.2$ from SDSS DR7 \citep{2009ApJS..182..543A}. The final catalog includes 1179 (494 detections out of 781 from GASS, and 310 detections out of 398 from GASS-low) with redshift range from 0.01 to 0.05.
        
        The xGASS stellar masses were from the improved version of the Max Planck Institute for Astrophysics (MPA)/Johns Hopkins University (JHU) value-added catalog based on SDSS DR7\footnote{\url{https://home.strw.leidenuniv.nl/~jarle/SDSS/}}. The xGASS SFRs were determined by combining the UV data from GALEX and MIR data from WISE \citep{2017MNRAS.466.4795J}. For consistency, as already noted, we adopted stellar masses and SFRs from GSWLC-2 for xGASS sources. Since the two catalogs use different SDSS catalog versions, we use the following criteria to cross-match the sources: 1) angular distance smaller than Petrosian radius in the $r$-band, and 2) recession velocity difference less than 200 km s$^{-1}$. Only seven xGASS galaxies do not have counterparts in the GSWLC-2 catalog. Ten counterparts in the GSWLC-2 catalog have no available stellar masses and SFRs. In total, there are 787 \HI\ detections from xGASS with stellar masses and SFRs available from GSWLC-2. These galaxies are used in the following comparisons.

\begin{figure}
            \begin{center}
                \includegraphics[width=\columnwidth]{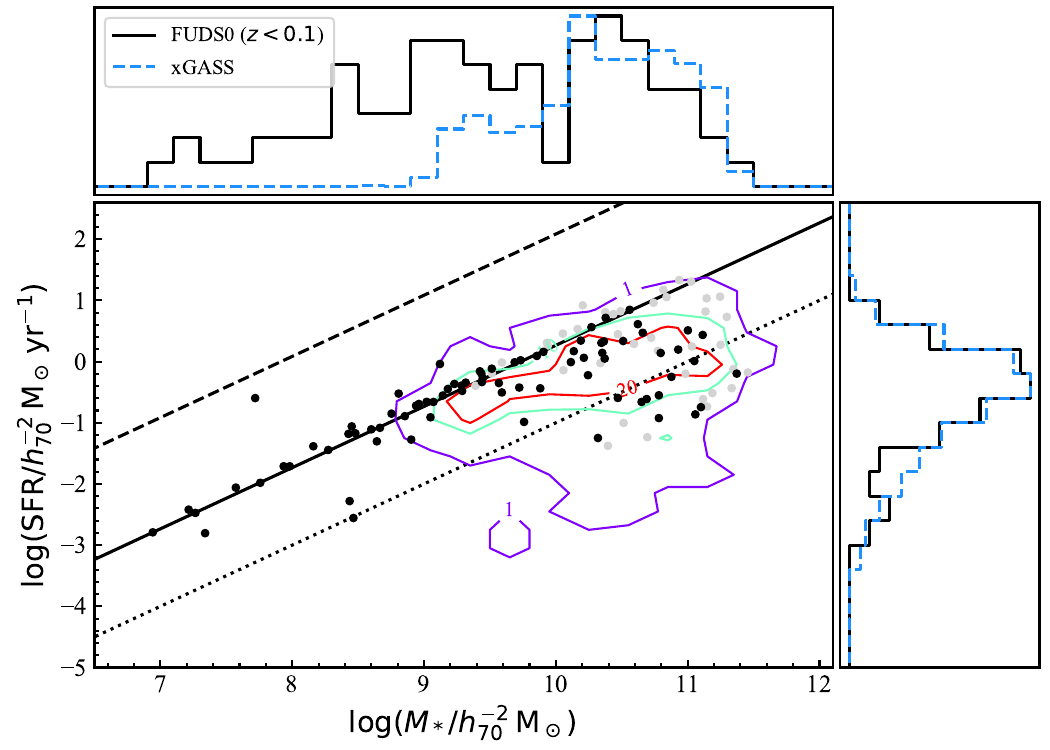}
                \caption{The comparison between FUDS0 and xGASS catalog. The black dots indicate the FUDS0 galaxies at $z<0.1$, while the light gray dots represents the rest of FUDS0 \HI\ galaxies. The black lines have same meanings as that in Figure \ref{Fig_15}. The contours represent the distribution of xGASS galaxies. The distributions of stellar mass (upper panel) and SFR (right panel) are normalized by the maximum.}\label{Fig_20}
            \end{center}
        \end{figure}

        Figure \ref{Fig_20} shows the comparison between the FUDS0 and xGASS catalogs. Since there are only 19 FUDS0 detections having stellar masses and SFRs in the xGASS redshift range ($z<0.05$), we extended the redshift to 0.1 for better comparison. There are 91 FUDS0 \HI\ galaxies at $z<0.1$, of which 79 have adjusted stellar masses and SFRs from \textsc{ProSpect}. 
        
        Both catalogs have a similar distribution in the overlapping stellar mass range ($\log(M_*/h_{70}^{-2} \rm M_\odot) \geq 9$). The histograms also show the consistency in SFR distribution, as well as in stellar mass distribution at the high-mass end. Although xGASS is an \HI\ survey of stellar mass-selected galaxies, it is designed to uniformly sample the galaxies across stellar mass. Therefore, xGASS will have less galaxies in the green valley and red sequence compared with other optically-selected catalogs such as GSWLC-2. FUDS0 has more low stellar mass galaxies at $\log(M_*/h_{70}^{-2} \rm M_\odot)<10$, due to the xGASS stellar mass cutoff, and  because \HI-selected galaxies generally have lower stellar mass than SDSS sources \citep{2010AJ....139..315W}.

        Figure \ref{Fig_20} also shows the main sequence line from the FUDS0 catalog (black solid line). xGASS galaxies have significantly lower SFR than the main sequence line, especially for $\log(M_*/h_{70}^{-2} \rm M_\odot) \geq 9.5$. A similar, but smaller offset is apparent for GSWLC-2 galaxies in the low redshift bin in Figure \ref{Fig_19}, and seems to be due to some Quenching stage galaxies (defined in Section \ref{Sct_06}) being classified as belonging to the SF sequence at high stellar mass.

    \subsection{HIGHz}

        The HIGHz Arecibo Survey (HIGHz; \citealp{2015MNRAS.446.3526C}) was aimed at investigating the \HI\ content in 49 optically selected sources with spectroscopic redshifts from SDSS (DR1, \citealp{2003AJ....126.2081A}, DR2, \citealp{2004AJ....128..502A}, DR3, \citealp{2005AJ....129.1755A}, DR5, \citealp{2007ApJS..172..634A}, DR7, \citealp{2009ApJS..182..543A} -- the latest version employed when the observations were performed) to study the evolution of \HI\ gas in galaxies. Considering the large beam size of the Arecibo telescope, the targets were deliberately selected to be isolated  to avoid confusion. As a result, 39 out of 49 galaxies were detected in \HI. The detections have stellar masses $\log(M_* / h_{70}^{-2} {\rm M_\odot}) > 10$, and redshifts $0.17 \leq z \leq 0.25$. 
        
        In the catalog, the stellar masses and SFRs were from the improved version of MPA/JHU SDSS DR7 catalog using the methodologies described in \citet{2007ApJS..173..267S} and \citet{2004MNRAS.351.1151B}, respectively. As before, we instead used stellar masses and SFRs from GSWLC-2 for consistency. Similar criteria to those for xGASS (see Section \ref{Sct_07_02}) were used to cross match HIGHz with GSWLC-2. 33 out of 39 HIGHz galaxies had GSWLC-2 counterparts. We use these galaxies as a non-local sample for comparisons.

\begin{figure}
            \begin{center}
                \includegraphics[width=\columnwidth]{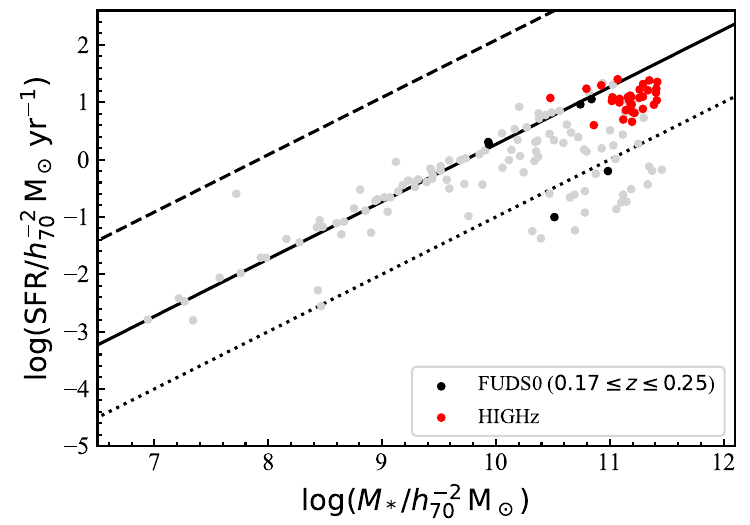}
                \caption{The comparison between FUDS0 and HIGHz catalog. The black dots indicate the FUDS0 galaxies in the same redshift range as HIGHz ($0.17 \leq z \leq 0.25$), while the light gray dots the rest of FUDS0 \HI\ galaxies. The black lines have same meanings as that in Figure \ref{Fig_15}. The red dots represent the HIGHz galaxies.}\label{Fig_21}
            \end{center}
        \end{figure}

        The HIGHz survey has a redshift range between 0.17 and 0.25 where there are only 10 FUDS0 galaxies, of which only 6 have adjusted stellar masses and SFRs from \textsc{ProSpect}. Figure \ref{Fig_21} shows that HIGHz galaxies. Although they have extreme stellar masses and SFRs, they still follow the SF sequence line defined by FUDS0 galaxies. Since one of the HIGHz selection criteria is the existence of an H$_\alpha$ emission line, the targets of HIGHz survey are automatically SF galaxies. Comparing HIGHz with xGASS, \citet{2015MNRAS.446.3526C} found that HIGHz galaxies are the analogues of the extreme galaxies in local universe, and found no evolutionary trend of \HI\ in galaxies in this redshift range.

    \subsection{ALFALFA}

        The Arecibo Legacy Fast ALFA (ALFALFA) survey \citep{2005AJ....130.2598G} is a blind \HI\ survey for studying \HI\ gas in nearby galaxies with 21 cm emission line. The Arecibo L-band Feed Array (ALFA) was employed to map over 7000 deg$^2$ in the SDSS footprint. 31,501 \HI\ galaxies were cataloged with redshifts between 0 and 0.06 \citep{2018ApJ...861...49H}. The SDSS counterparts of 29,638 galaxies were identified \citep{2020AJ....160..271D}. Three methods were employed to derive stellar mass: $M_{*, \rm Taylor}$ from the SDSS $g$- and $i$-band flux densities  \citep{2011MNRAS.418.1587T}; $M_{*, \rm McGaugh}$ from the WISE $W1$ band \citep{2015ApJ...802...18M}; and $M_{*, \rm Cluver}$ from the WISE $W1$ and $W2$ bands \citep{2014ApJ...782...90C}. However, $M_{*, \rm Cluver}$ has a larger scatter and offset for low stellar mass galaxies ($\log(M_*/h_{70}^{-2} {\rm M_\odot})<10$). The SFRs were also calculated using three methods: SFR$_{22}$ from the WISE $W4$ band 22 $\mu$m flux densities  \citep{2012ARA&A..50..531K}; SFR$_{\rm NUV}$ from the GALEX NUV data \citep{2011ApJ...741..124H, 2011ApJ...737...67M, 2012ARA&A..50..531K}; and SFR$_{\rm NUV_{\rm Corr}}$ from GALEX NUV data corrected with 22 $\mu$m flux densities from WISE \citep{2011ApJ...741..124H, 2012ARA&A..50..531K}.

        Unlike for the xGASS and HIGHz catalogs, there is only a low fraction ($\sim 40\%$) of ALFALFA sources identified in the GSWLC-2 catalog \citep{2020AJ....160..271D}. Hence, we use GSWLC-2 as a reference to adjust stellar masses and SFRs for ALFALFA sources, as we did for FUDS0. Fortunately, these comparisons are already available in \citet{2020AJ....160..271D}. Based on the scatter and offset, the authors gave a priority for each method as follows: $P(M_{*, \rm Taylor}) > P(M_{*, \rm McGaugh}) > P(M_{*, \rm Cluver})$, and $P({\rm SFR}_{\rm NUV_{\rm Corr}}) > P({\rm SFR}_{\rm NUV}) > P({\rm SFR}_{22})$. We also adopted the same priority order. The adjustments were performed by using the best fit lines of \citet{2020AJ....160..271D}, given below (note that the paper does not provide $M_{*, \rm Cluver}$ or ${\rm SFR}_{\rm NUV}$): 
        \begin{equation}
            \begin{split}
                & \log(M_{*}/h_{70}^{-2} {\rm M_\odot}) \\
                & = 1.052 \cdot \log(M_{*, \rm Taylor}/h_{70}^{-2} \rm M_\odot)-0.369
            \end{split}
        \end{equation}
        
        \begin{equation}
            \begin{split}
                & \log(M_{*}/h_{70}^{-2} {\rm M_\odot}) \\
                & = 1.084 \cdot \log(M_{*, \rm McGaugh}/h_{70}^{-2} \rm M_\odot)-0.9755
            \end{split}
        \end{equation}

        \begin{equation}
            \begin{split}
                & \log({\rm SFR}/h_{70}^{-2} {\rm M_\odot \, yr^{-1}}) \\
                & = \log({\rm SFR}_{22}/h_{70}^{-2} {\rm M_\odot \, yr^{-1}})+0.09
            \end{split}
        \end{equation}

        \begin{equation}
            \begin{split}
                & \log({\rm SFR}/h_{70}^{-2} {\rm M_\odot \, yr^{-1}}) \\
                & = \log({\rm SFR}_{\rm NUV_{\rm Corr}}/h_{70}^{-2} {\rm M_\odot \, yr^{-1}})-0.09 .
            \end{split}
        \end{equation}
        This resulted in 29,568 ALFALFA sources with adjusted stellar masses, 23,895 with adjusted SFRs, and 23,869 with both. The ALFALFA sources with both adjusted stellar mass and SFR are used for further comparisons.

\begin{figure}
            \begin{center}
                \includegraphics[width=\columnwidth]{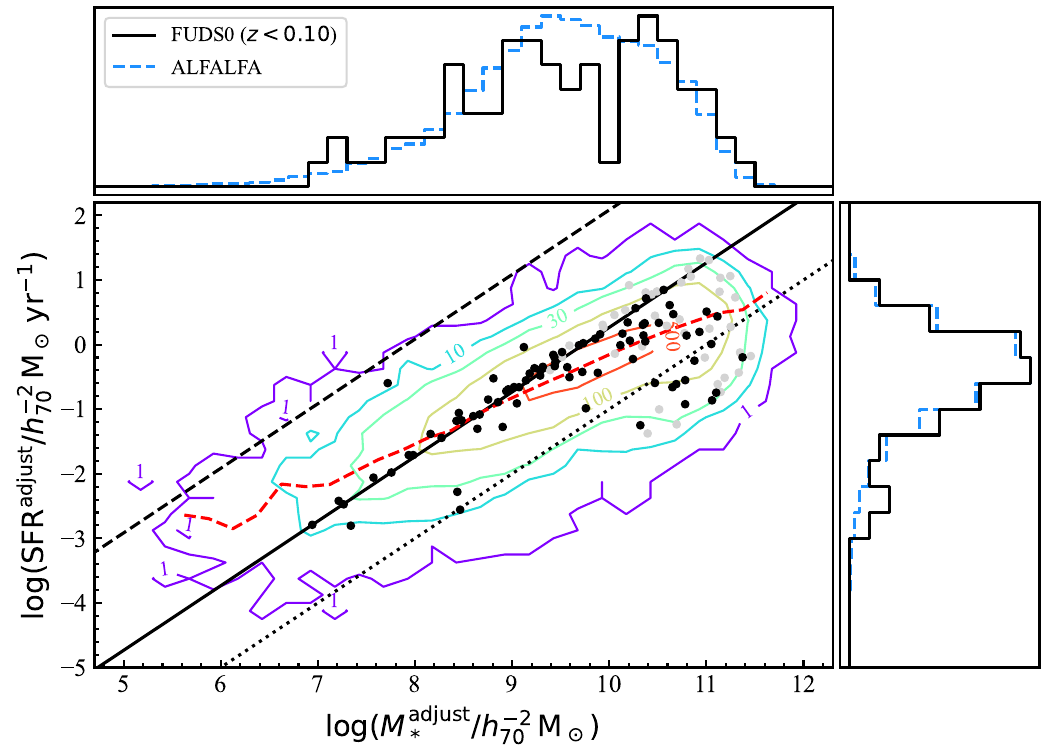}
                \caption{The comparison between the FUDS0 and ALFALFA catalogs. The black dots represent the FUDS0 galaxies located at $z<0.1$, while the gray dots others. The contours display the distribution of ALFALFA galaxies. The red dashed line indicates the median ALFALFA SFR versus stellar mass. The black lines are the same as for Figure \ref{Fig_15}.}\label{Fig_22}
            \end{center}
        \end{figure}

        ALFALFA sources have redshifts $z<0.06$, where there are only 29 FUDS0 \HI\ galaxies with adjusted stellar masses and SFRs from \textsc{ProSpect}. For a better comparison, the redshift limit was again extended to 0.1, where there are a total of 79 FUDS0 detections having adjusted stellar masses and SFRs. Figure \ref{Fig_22} shows the distribution of ALFALFA and FUDS0 galaxies in the SFR versus stellar mass diagram. 
        
        The histograms reveal almost identical distributions in both stellar mass and SFR. We do not find any obvious discrepancies at either mass end, as seen with GSWLC-2 and xGASS. The consistency also shows that there is no obvious bias in the stellar mass and SFR for FUDS0 galaxies, although the sky coverage is small.
        
        In the stellar mass -- SFR plane, FUDS0 galaxies follow the ridge of the distribution of ALFALFA galaxies at $\log(M_*/h_{70}^{-2} \rm M_\odot) > 7.5$. These galaxies are also located close to the median SFR line (dashed red line) of ALFALFA galaxies. Again, the SF sequence line from FUDS0 \HI\ galaxies is above the median SFR line from ALFALFA above $\log(M_*/h_{70}^{-2} \rm M_\odot) = 9.5$. This difference is similar to the comparisons with GSWLC-2 or xGASS, and we attribute it to interlopers from the Quenching stage (defined in Section \ref{Sct_06}). At the low-mass end, the FUDS0 galaxies lie below the ALFALFA median SFR. The offset is similar to that seen with GSWLC-2. This is due to the inclusion of Early stage galaxies (defined in Section \ref{Sct_06}), classified by the sSFR lower limit. There are only a few Early stage galaxies (defined in Section \ref{Sct_06}) with $\log(M_*/h_{70}^{-2} \rm M_\odot) > 7.5$ and they have little impact on the calculation of median SFR. However, they dominate the low mass end, where the number of Mass accumulation stage galaxies (defined in Section \ref{Sct_06}) is comparable. We can also see the trend in the comparison between FUDS0 and GSWLC-2. However, as the GSWLC-2 catalog does not consist of any galaxies with $\log(M_*/h_{70}^{-2} \rm M_\odot) < 7.5$, the discrepancy is not as clear. 

\section{Summary}\label{Sct_08}

    In this paper, we present the counterparts of FUDS0 galaxies in the UV, optical, and IR bands. Their stellar masses, SFRs and SFHs are derived by SED fitting and comparisons are made with previous work. The main findings are summarized below:

    \begin{enumerate}
    
        \item The counterparts of 128 FUDS0 galaxies include 118 from the DESI Imaging Legacy Survey, 117 from SDSS, 117 from unWISE and 97 from GALEX. There are 62 galaxies with optical spectroscopic redshifts, while the remaining 66 have only photometric redshfits.

        \item There are 134 galaxies in the FUDS0 field with optical spectroscopic redshifts, but not detected in \HI. Upper limits to their \HI\ masses are derived.

        \item Stellar masses and SFRs as functions of time are derived for the 117 FUDS0 \HI\ galaxies and the 134 non-detections by SED fitting using \textsc{ProSpect}. The peak sSFR is $\log({\rm sSFR / yr^{-1}}) = -7.9$ at any time in the SFH of FUDS0 galaxies. The current sSFR for SF sequence galaxies is $\log({\rm sSFR / yr^{-1}}) = -9.73 \pm 0.18$, except for high-mass galaxies, where it is lower.

        \item The \textsc{ProSpect} evolutionary tracks of the \HI\ galaxies in the SFR versus stellar mass plane suggest the following typical evolutionary scenario: ({\romannumeral 1}) an Early stage, where SFR increases as stellar mass accumulates, and the sSFR drops from $10^{-8}$ to $10^{-9}$ yr$^{-1}$; ({\romannumeral 2}) a Mass accumulation stage, where SFR is nearly constant and the stellar mass increases linearly with time; and ({\romannumeral 3}) a Quenching stage, where the SFR drops dramatically with stellar mass remaining nearly constant.

        \item Comparisons of various subsets of FUDS0 with existing catalogs, including GSWLC-2, xGASS, HIGHz and ALFALFA, show a good agreement in stellar masses and SFRs, with minor differences which appear due to selection effects, and support the evolutionary scenario proposed here.
        
    \end{enumerate}

    FUDS0 is the pilot survey for FUDS. The full FUDS catalog will provide a six times larger sample with lower cosmic variance and will better explore the properties and evolution of \HI\ galaxies. The data quality of subsequent observations will also be better due to the elimination of internal sources of RFI.

%

\section{Acknowledgements}

    We would like to thank Aaron Robotham for his assistance with \textsc{ProSpect}. This work made use of the data from FAST (Five-hundred-meter Aperture Spherical radio Telescope)(\url{https://cstr.cn/31116.02.FAST}). FAST is a Chinese national mega-science facility, operated by National Astronomical Observatories, Chinese Academy of Sciences. The work is supported by the National Key R\&D Program of China under grant number 2018YFA0404703, the Science and Technology Innovation Program of Hunan Province under grant number 2024JC0001, and the FAST Collaboration.

    The DESI Legacy Imaging Surveys consist of three individual and complementary projects: the Dark Energy Camera Legacy Survey (DECaLS), the Beijing-Arizona Sky Survey (BASS), and the Mayall z-band Legacy Survey (MzLS). DECaLS, BASS and MzLS together include data obtained, respectively, at the Blanco telescope, Cerro Tololo Inter-American Observatory, NSF’s NOIRLab; the Bok telescope, Steward Observatory, University of Arizona; and the Mayall telescope, Kitt Peak National Observatory, NOIRLab. NOIRLab is operated by the Association of Universities for Research in Astronomy (AURA) under a cooperative agreement with the National Science Foundation. Pipeline processing and analyses of the data were supported by NOIRLab and the Lawrence Berkeley National Laboratory (LBNL). Legacy Surveys also uses data products from the Near-Earth Object Wide-field Infrared Survey Explorer (NEOWISE), a project of the Jet Propulsion Laboratory/California Institute of Technology, funded by the National Aeronautics and Space Administration. Legacy Surveys was supported by: the Director, Office of Science, Office of High Energy Physics of the U.S. Department of Energy; the National Energy Research Scientific Computing Center, a DOE Office of Science User Facility; the U.S. National Science Foundation, Division of Astronomical Sciences; the National Astronomical Observatories of China, the Chinese Academy of Sciences and the Chinese National Natural Science Foundation. LBNL is managed by the Regents of the University of California under contract to the U.S. Department of Energy. The complete acknowledgments can be found at \url{https://www.legacysurvey.org/acknowledgment/}.

    The Photometric Redshifts for the Legacy Surveys (PRLS) catalog used in this paper was produced thanks to funding from the U.S. Department of Energy Office of Science, Office of High Energy Physics via grant DE-SC0007914.

%

\vspace{5mm}
\facilities{FAST, Hale, BTA, Keck:I, GALEX, Sloan, WISE}


\software{\textsc{astropy} \citep{2013A&A...558A..33A,2018AJ....156..123A, 2022ApJ...935..167A}, \textsc{matplotlib} \citep{4160265}, \textsc{NumPy} \citep{5725236}, \textsc{ProSpect} \citep{2020MNRAS.495..905R}}, \textsc{dustmaps} \citep{2018JOSS....3..695G}, \textsc{extinction}



\appendix

\section{Impact of noise}\label{Sct_A}

    FUDS is a deep survey with a high sensitivity ($\sim 50 \, \mu$Jy at the center of the field). However, unlike HIPASS and ALFALFA, the sampling time is not uniformly distributed spatially (due to edge effects) or spectrally (due to RFI), so the local noise values are not constant. In this Section, we evaluate the impact of noise on the distribution of FUDS0 galaxies and the effect on comparison with previous work.

\begin{figure}
        \begin{center}
            \includegraphics[width=\columnwidth]{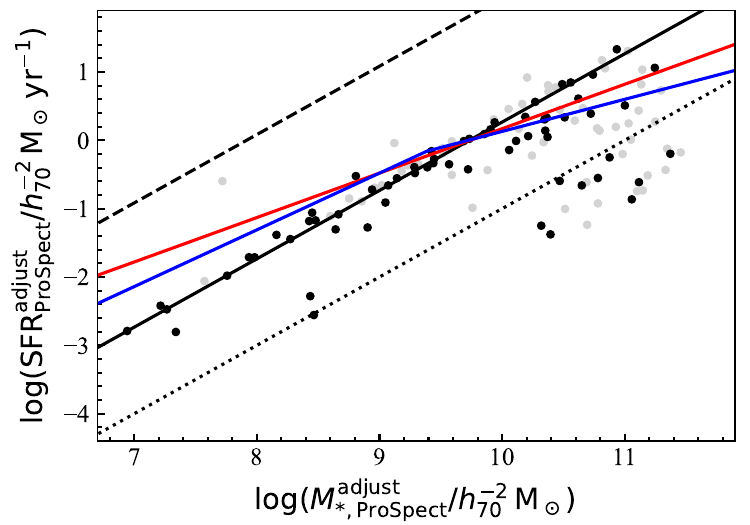}
            \caption{The distribution of FUDS0 galaxies in the SFR -- $M_*$ plane. The black and gray dots represent strong or weak detections separated by $T_{\sigma}=\frac{S_{\rm int}}{\rm Jy Hz} (\frac{\rm 10^6 Hz}{W_{20}})^\alpha = 250$. The black lines have same meanings as that in Figure \ref{Fig_15}}\label{Fig_23}
        \end{center}
    \end{figure}

    In \citet{2024ApJS..274...18X}, we analyzed the detectability (completeness) and its dependencies, and found that $T=\frac{S_{\rm int}}{\rm Jy Hz} \frac{\rm Jy}{\sigma} (\frac{\rm 10^6 Hz}{W_{20}})^\alpha$ is a useful detection threshold. Therefore, for a survey with uniform noise, $T_{\sigma}=\frac{S_{\rm int}}{\rm Jy Hz} (\frac{\rm 10^6 Hz}{W_{20}})^\alpha$ is an ideal indicator to distinguish if a galaxy can be detected. Here we defined two virtual surveys, `FUDS0 shallow' ($T_{\sigma}>250$) and `FUDS0 deep' ($T_{\sigma} \leq 250$). There are 64 galaxies in `FUDS0 shallow' and 64 galaxies in `FUDS0 deep'. The distribution in the SFR -- $M_*$ plane of `FUDS0 shallow' and `FUDS0 deep' galaxies is shown in Figure \ref{Fig_23}. Most of the galaxies from both virtual surveys follows the same SF sequence line (black line). This consistency suggests that there are no particular biases which will affect the comparison between FUDS0 and other \HI\ surveys.

\begin{figure}
        \begin{center}
            \includegraphics[width=\columnwidth]{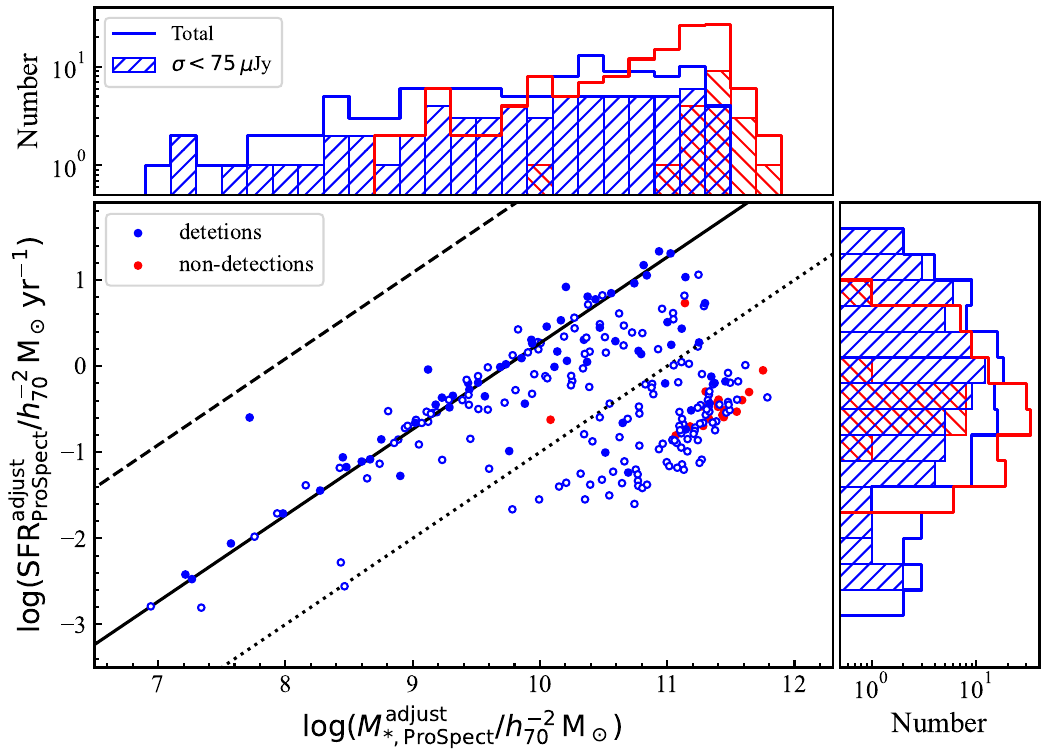}
            \caption{The distribution of galaxies in the SFR -- $M_*$ plane. The galaxies with local noise $\sigma < 75 \, \mu$Jy are represented by filled dots, while the others by empty dots. The blue dots indicates the FUDS0 \HI\ galaxies, while red dotes non-detections. The black lines are the same as for Figure \ref{Fig_15}. The distribution of stellar mass (upper panel) and SFR (right-hand side panel) are given by the hatched histogram with same color. The solid lines indicate the full sample.}\label{Fig_24}
        \end{center}
    \end{figure}

    We also define a FUDS0 `High sensitivity' subset where the local noise $\sigma < 75 \, \mu$Jy. There are 74 FUDS0 \HI\ galaxies and 19 non-detections in this subset, distributed in the SFR -- $M_*$ as shown in Figure \ref{Fig_24}. The \HI\ galaxies in the subset has a similar distribution to the full FUDS0 \HI\ galaxy sample, which also indicates there will be little impact of noise variation on the full FUDS0 \HI\ galaxy catalog for comparisons with other catalogs. Considering that the SF sequence is dominated by \HI\ galaxies, it is possible that confusion may be responsible for some \HI\ galaxies having low SFR \citep{2015MNRAS.451..103Y}. However, the distribution of non-detections is different from that in full FUDS0 non-detection sample. Most of these are galaxies with high stellar mass and low SFR (green valley and red sequence), which suggest early-type galaxies. The difference in distribution could indicate inclusion of \HI\ galaxies. In the full sample, there are 37 non-detections with $\sigma \geq 75 \, \mu$Jy in the SF sequence ($\log({\rm sSFR}/{\rm yr^{-1}}) > -11$). However, these non-detections will probably be \HI\ galaxies that are missed by FUDS0 due to low sensitivity at frequencies near RFI or regions close to the field edge.

\section{Evolution snapshots}\label{Sct_B}

\begin{figure*}
        \begin{center}
            \includegraphics[width=0.8\columnwidth]{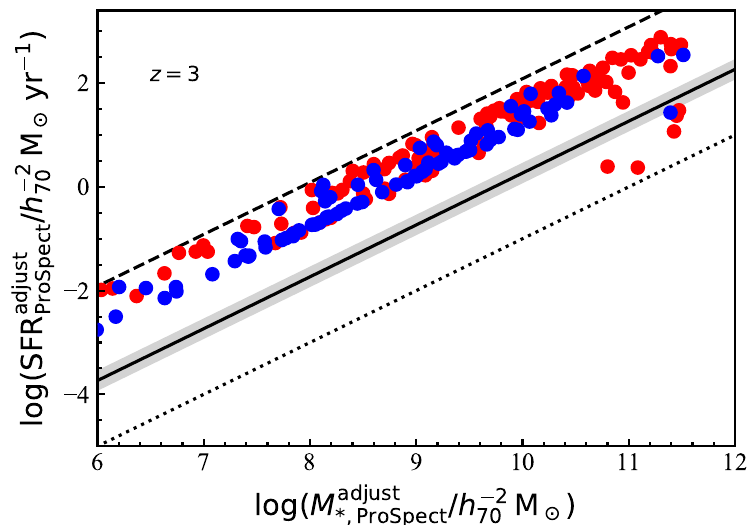}
            \includegraphics[width=0.8\columnwidth]{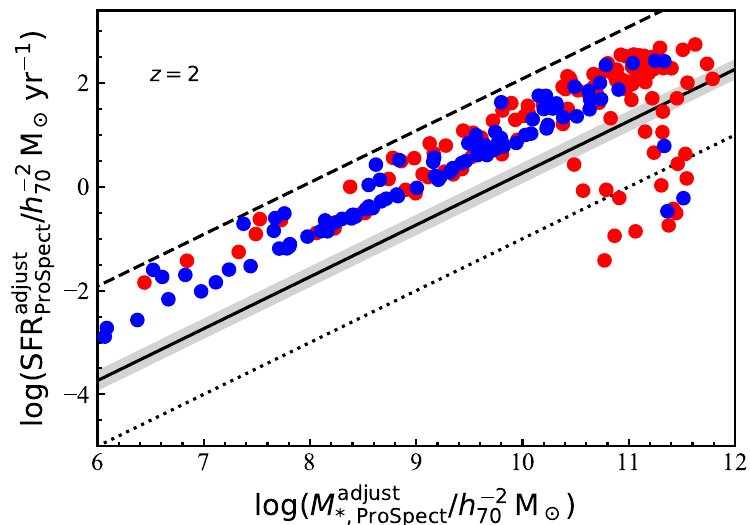}
            \includegraphics[width=0.8\columnwidth]{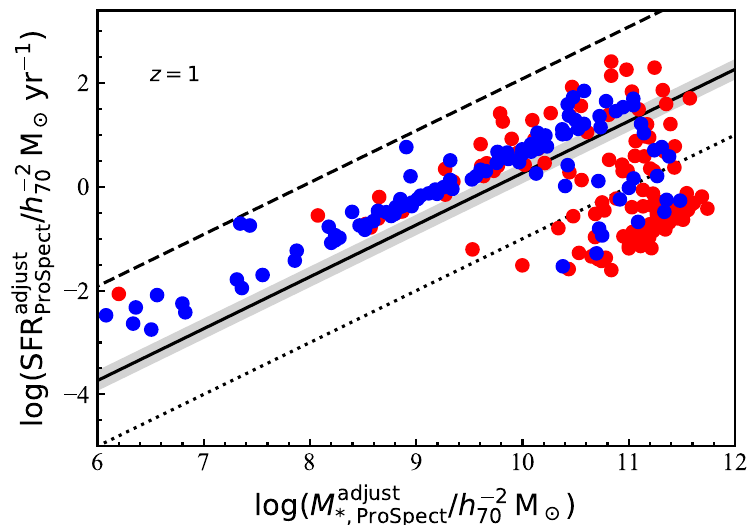}
            \includegraphics[width=0.8\columnwidth]{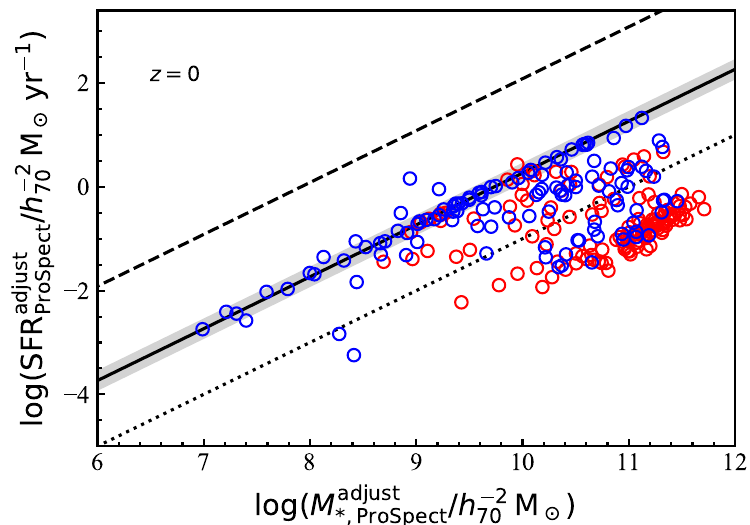}
            \caption{The distributions of both detections (blue) and non-detections (red) in the SFR -- $M_*$ plane at $z = 3, 2, 1, 0$ as predicted by \textsc{ProSpect}. The filled circles indicate that the redshift is larger than the observed redshift of the galaxies, while the empty circles indicate where the redshift is smaller than the observed redshift of the galaxies. The black lines have the same meanings as that in Figure \ref{Fig_15}.}\label{Fig_25}
        \end{center}
    \end{figure*}

\begin{figure*}
        \begin{center}
            \includegraphics[width=0.8\columnwidth]{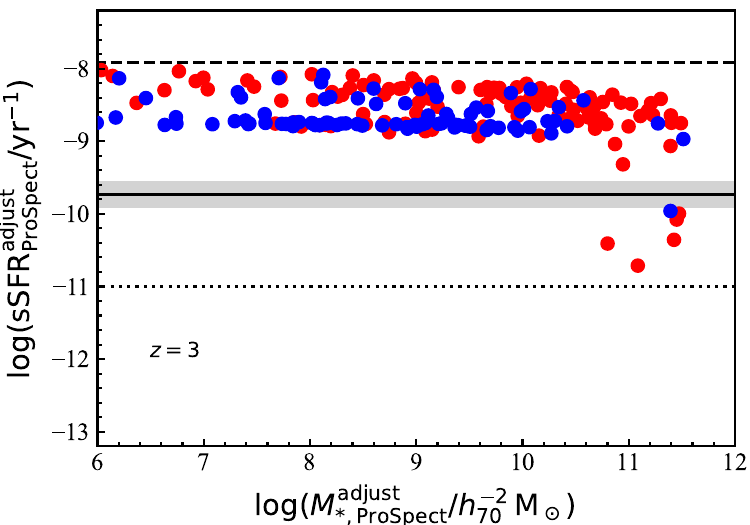}
            \includegraphics[width=0.8\columnwidth]{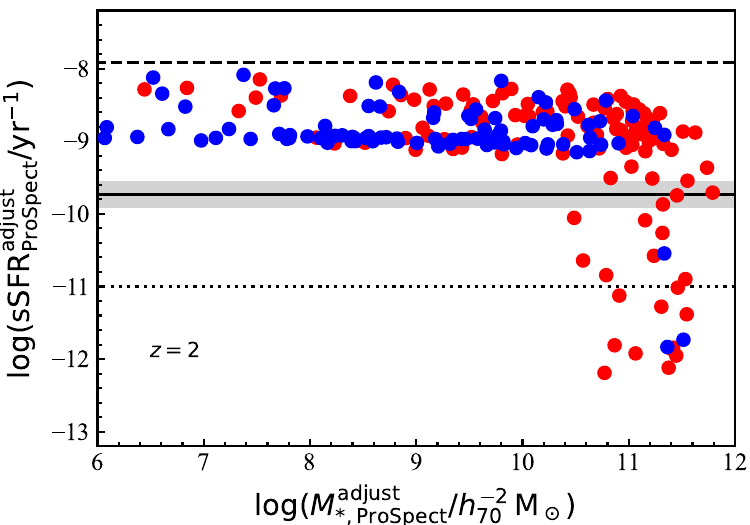}
            \includegraphics[width=0.8\columnwidth]{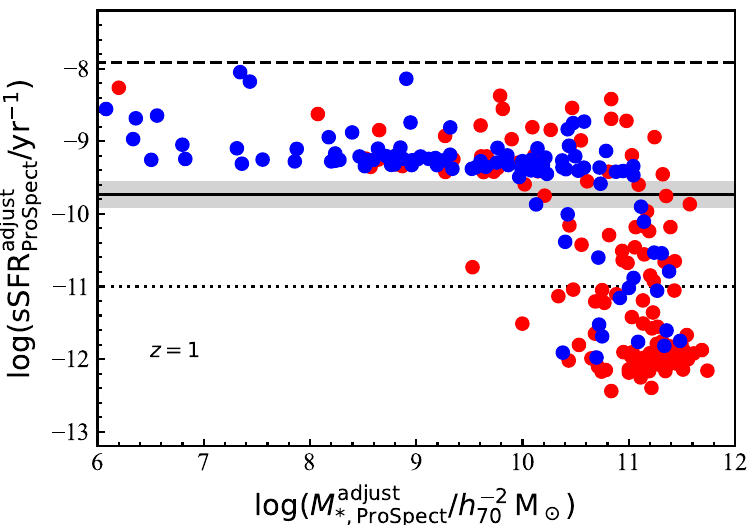}
            \includegraphics[width=0.8\columnwidth]{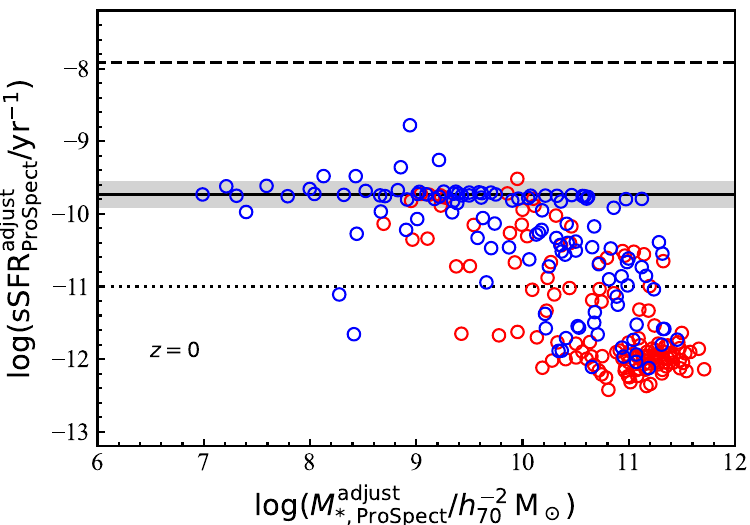}
            \caption{The distributions of both detections (blue) and non-detections (red) in the sSFR -- $M_*$ plane at $z = 3, 2, 1, 0$ as predicted by \textsc{ProSpect}. The filled and empty circles have same meaning as those in Figure \ref{Fig_25}. The black lines have the same meanings as that in Figure \ref{Fig_15}.}\label{Fig_26}
        \end{center}
    \end{figure*}

    For a more detailed illustration of the evolution of \HI\ galaxies and non-detections,  snapshots of their physical properties are generated at different redshifts ($z=3, 2, 1, 0$). The tracks of FUDS0 galaxies and non-detections in SFR--$M_*$ and sSFR--$M_*$ planes are shown in Figures \ref{Fig_25} and \ref{Fig_26}, respectively. The three evolutionary stages for \HI\ galaxies are clearly shown in the figures. Different evolutionary tracks for the non-detections can also be seen. These spend a shorter period in the Mass accumulation stage (defined in Section \ref{Sct_06}).
    
\bibliography{References}{}
\bibliographystyle{aasjournal}



\end{document}